\documentclass[a4paper,11pt]{article}
\pdfoutput=1 

\usepackage{jheppub} 
\usepackage{grffile} 
\usepackage{slashed,ulem,color}

\newcommand{\tr}{\mbox{tr}}
\newcommand{\VEV}[1]{\langle #1 \rangle}

\title{A model of electroweakly interacting non-abelian vector dark matter}
\author[a,b]{Tomohiro Abe}
\author[c]{Motoko Fujiwara}
\author[b,c,d]{Junji Hisano}
\author[c]{Kohei Matsushita}

\affiliation[a]{
  Institute for Advanced Research, Nagoya University,
  Furo-cho Chikusa-ku, Nagoya, Aichi, 464-8602 Japan
}
\affiliation[b]{
  Kobayashi-Maskawa Institute for the Origin of Particles and the
  Universe, Nagoya University,
  Furo-cho Chikusa-ku, Nagoya, 464-8602 Japan
}
\affiliation[c]{
  Department of Physics, Nagoya University, Furo-cho Chikusa-ku, Nagoya, 464-8602 Japan
}
\affiliation[d]{
  Kavli IPMU (WPI), UTIAS, University of Tokyo, Kashiwa, 277-8584, Japan
}
\emailAdd{abetomo@kmi.nagoya-u.ac.jp}
\emailAdd{motoko@eken.phys.nagoya-u.ac.jp}
\emailAdd{hisano@eken.phys.nagoya-u.ac.jp}
\emailAdd{kohei@eken.phys.nagoya-u.ac.jp}

\abstract{
We propose an electroweakly interacting spin-1 dark matter (DM) model. 
The electroweak gauge symmetry,
SU(2)$_L\times$U(1)$_Y$, is extended into  
SU(2)$_0\times$SU(2)$_1 \times $SU(2)$_2 \times$U(1)$_Y$.
A discrete symmetry exchanging  SU(2)$_0$ and SU(2)$_2$ is imposed. This discrete symmetry stabilizes the DM candidate.
The spin-1 DM particle ($V^0)$ and its SU(2)$_L$ partners ($V^\pm$) interact with the Standard Model (SM) electroweak gauge bosons without any suppression factors. 
Consequently, pairs of DM particles efficiently annihilate into the SM particles in the early universe, 
and the measured value of the DM energy density is easily realized by the thermal freeze-out mechanism.
The model also predicts a heavy vector triplet ($W'^\pm$ and $Z'$) in the visible sector. 
They contribute to the DM annihilation processes. The mass ratio of $Z'$ and $V^0$ determines values of various couplings, and constraints on $W'$ and $Z'$ restrict regions of the parameter space that are viable for DM physics.
We investigate the constraints from
perturbative unitarity of scalar and gauge couplings, the Higgs signal strength, $W'$ search at the LHC, and DM direct detection experiments.
It is found that the relic abundance of $V^0$ explains the right amount of the DM energy density for 3~TeV $\lesssim m_{V^0} \lesssim$ 19~TeV.
}
\keywords{}
\arxivnumber{2004.00884}

\begin{document}

\maketitle


\section{Introduction}

Dark matter (DM) is a longstanding issue in both particle physics and cosmology. 
The DM energy density is precisely measured by the Planck collaboration, $\Omega h^2 = 0.120\pm 0.001$~\cite{1807.06209}.
A popular scenario that explains this measured value is the thermal freeze-out scenario~\cite{Lee:1977ua}, which utilizes a pair annihilation/creation of DM particles into/from particles in the thermal bath in the early universe.
This scenario requires interactions between DM and Standard Model (SM) particles.
Hence, models that utilize the freeze-out mechanism are good targets of DM direct detection experiments. However, there are no significant DM signals at the experiments so far. The latest result by the XENON1T experiment gives a strong upper bound on the DM-nucleon scattering cross section~\cite{1805.12562}. 
This result implies that the DM-nucleon scattering processes mediated by the $Z$-boson and scalar mediators, such as the SM Higgs boson, must be suppressed if the mass of DM is between ${\cal O}$(10)~GeV and ${\cal O}$(1)~TeV. 

Models that predict the suppression in those processes without suppressed DM-mediator coupling are proposed. Fermionic DM models with pseudo-scalar couplings are examples in spin-1/2 DM models~\cite{1404.3716,1408.4929, 1701.04131}.
The pseudo-Nambu-Goldstone (NG) boson DM models are examples of spin-0 DM models~\cite{1708.02253, Abe:2020iph, Okada:2020zxo, Ahmed:2020hiw}.
In these models, the coupling itself between DM and mediator particle is not suppressed, and thus DM is thermally produced in the early universe 
through the annihilation into the SM particles via scalar mediator exchanges.

In this paper, we propose a renormalizable model of spin-1 DM that does not require $Z$ and Higgs couplings to a DM particle to obtain the correct amount of the DM density by the freeze-out mechanism.\footnote{
Non-renormalizable models for the electroweakly interacting spin-1 DM are discussed in~\cite{Maru:2018ocf, Belyaev:2018xpf}.
} We extend the electroweak gauge symmetry in the SM, SU(2)$_L \times$U(1)$_Y$, into SU(2)$_0 \times $SU(2)$_1 \times$SU(2)$_2 \times$U(1)$_Y$ and impose that the model is symmetric under exchanging of SU(2)$_0$ and SU(2)$_2$. 
This symmetry predicts a stable SU(2)$_L$ triplet vector boson, $V^0$ and $V^\pm$. After the symmetry breaking, the charged vector boson, $V^\pm$, gets slightly heavier than the neutral one, $V^0$, and thus $V^0$ is a DM candidate in our model.
The vector DM in our model can directly couple to the SM weak gauge bosons and efficiently annihilate in the early universe even without the DM-Higgs coupling. The $V^0$-$V^0$-$Z$ coupling is  automatically forbidden by the gauge symmetry. Therefore, the model easily evades the constraint from the XENON1T experiment and has a large region of viable parameter space.

There are many spin-1 DM models, but they are originated from a U(1) gauge symmetry~\cite{hep-ph/0206071, 1005.5651, 1111.4482, 1202.5902, 1207.4272, 1212.2131,1312.4573, 1404.5257, 1405.3530, 1409.3227, 1410.0918} 
 or an  SU(2) gauge symmetry that is isolated from the SM electroweak sector~\cite{0811.0172, 0910.2831, 1007.2631,1107.2093,1306.2329, 1309.6640,
 Baek:2013dwa, 1406.2291, 1409.1162, Karam:2015jta}.
 Therefore, they rely on the scalar exchanges that require the mixing between the SM Higgs and new scalar particles to obtain the measured value of the DM energy density.
The scalar mixing, however, is constrained from the direct detection experiments. On the other hand, our model does not require the scalar mixing for the DM energy density. This is a different feature of our model from the other spin-1 DM models.
Another aspect of our model is that new spin-1 particles are predicted in the visible sector as well as the dark sector. Those new spin-1 particles in the visible sector are regarded as $W'$ and $Z'$. They play an important role in the DM annihilation processes. 
Moreover, the fermion sector of our model is as simple as in the SM.
We do not need to introduce new fermions into the model to obtain the realistic mass spectra for the SM fermions.\footnote{Non-abelian vector DM with an extended fermion sector are discussed in~\cite{DiazCruz:2010dc, Barman:2017yzr, Barman:2018esi,Barman:2019lvm}.}

We organize the rest of this paper as follows.
In Sec.~\ref{sec:model}, we describe our model. 
Some technical details are discussed in Appendices.
In Sec.~\ref{sec:constraints}, we discuss constraints on the model from 
perturbative unitarity, the mass ratio of $Z'$ and $V^0$, $W'$ and $Z'$ searches at the LHC, electroweak precision measurements, and the Higgs coupling measurements at the LHC.
After constraining the model parameters, 
we discuss the phenomenology of DM in Sec.~\ref{sec:DM-pheno}.
We start by discussing the mass difference between $V^\pm$ and $V^0$. 
As discussed later, $V^\pm$ is one of the targets for  long-lived particle searches at the LHC.
After that, we discuss the thermal relic abundance in this model. 
We also address the constraint from the XENON1T experiment.
We show that the viable mass range of $V^0$ as a thermal relic is 3~TeV $\lesssim m_{V^0} \lesssim$19~TeV.
Section~\ref{sec:summary} is devoted to our conclusions.

\section{Model}\label{sec:model}
The gauge symmetry is SU(3)$_c \times$SU(2)$_0 \times$SU(2)$_1 \times$SU(2)$_2 \times$U(1)$_Y$ in our Model. Here, SU(3)$_c$ is for the QCD as in the same as the SM.
The matter and Higgs fields are summarized in Tab.~\ref{tab:matter}.\footnote{
A model with a similar gauge group is studied in~\cite{1712.08994} but with different matter contents and with different gauge charge assignments.
}
In this section, we focus on the extended electroweak gauge sector, namely SU(2)$_0 \times$SU(2)$_1 \times$SU(2)$_2 \times$U(1)$_Y$.
We denote the gauge fields of them as 
$W_{0\mu}^a$, $W_{1\mu}^a$, $W_{2\mu}^a$, and $B_{\mu}$, respectively, where $a = 1,2,3$.
Their gauge couplings are $g_0$, $g_1$, $g_2$, and $g'$, respectively.
The gauge transformation of two Higgs fields, $\Phi_1$ and $\Phi_2$ , are given by
\begin{align}
 \Phi_1 \to U_0 \Phi_1 U_1^\dagger, \\
 \Phi_2 \to U_2 \Phi_2 U_1^\dagger, 
\end{align}
where $U_j$'s are two-by-two unitary matrices of the SU(2)$_j$ gauge transformation.
To reduce the number of degrees of freedom, we impose 
\begin{align}
 \Phi_j = -\epsilon \Phi_j^* \epsilon, \quad
\text{where} \quad \epsilon = \begin{pmatrix} 0 & 1 \\ -1 & 0 \end{pmatrix}.
\end{align}
Before imposing this constraint, $\Phi_1$ and $\Phi_2$ contain four complex degrees of freedom (eight real degrees of freedom), respectively. After imposing this constraint, each field has four real degrees of freedom as shown later in Eq.~\eqref{eq:component_fields}.
This constraint has nothing to do with the dark matter stability.

We impose the following discrete symmetry.
\begin{align}
  &q_L  \to q_L, \quad u_R   \to u_R, \quad  d_R   \to d_R, \\
  &\ell_L \to \ell_L,\quad  e_R    \to e_R,\\
  &H      \to H,\quad 
  \Phi_1 \to \Phi_2,\quad
  \Phi_2 \to \Phi_1,\\
  &W_{0\mu}^a \to   W_{2\mu}^a,\quad
  W_{1\mu}^a \to   W_{1\mu}^a,\quad
  W_{2\mu}^a \to   W_{0\mu}^a.
\end{align}
This discrete symmetry is equivalent to the exchange of SU(2)$_0$ and SU(2)$_2$.
It requires $g_0 = g_2$.
The symmetry works as a $Z_2$ symmetry that is utilized in many dark matter models. 
Linear combinations $(W^{a}_{0\mu} - W^{a}_{2\mu})/\sqrt{2}$ are 
odd under the symmetry. They are mass eigenstates as we will see below, and one of them is  a DM candidate. On the other hand, the other linear combinations of the gauge fields are even under the symmetry.
Similarly, linear combinations of $\Phi_1$ and $\Phi_2$ divide scalar fields into the odd and even sectors.
All the SM particles are even under the discrete symmetry.
\begin{table}[tb]
\centering
\caption{The matter and Higgs fields and their gauge charges in the model. The generation indices for the matter fields are implicit.}
\label{tab:matter}
 \begin{tabular}{cc|ccccc}\hline
  field    & spin           & SU(3)$_c$ & SU(2)$_0$ & SU(2)$_1$ & SU(2)$_2$ & U(1)$_Y$ \\ \hline \hline
  $q_L$    & $\frac{1}{2}$  &   3   &   1       &    2      &     1     &   $\frac{1}{6}$ \\
  $u_R$    & $\frac{1}{2}$  &   3   &   1       &    1      &     1     &   $\frac{2}{3}$ \\
  $d_R$    & $\frac{1}{2}$  &   3   &   1       &    1      &     1     &  -$\frac{1}{3}$ \\
  $\ell_L$ & $\frac{1}{2}$  &   1   &   1       &    2      &     1     &  -$\frac{1}{2}$ \\
  $e_R$    & $\frac{1}{2}$  &   1   &   1       &    1      &     1     &  -1 \\ \hline
  $H$      & 0              &   1   &   1       &    2      &     1     &   $\frac{1}{2}$ \\
  $\Phi_1$ & 0              &   1   &   2       &    2      &     1     &   0 \\
  $\Phi_2$ & 0              &   1   &   1       &    2      &     2     &   0 \\ \hline \hline
 \end{tabular}
\end{table}

The discrete symmetry under exchanging SU(2)$_0$ and SU(2)$_2$ is inspired by the deconstruction~\cite{Hill:2000mu, ArkaniHamed:2001ca} of models in extra dimension on $S^1/Z_2$.
Using the deconstruction approach, such models are expressed by moose diagrams~\cite{Georgi:1985hf}.
The $Z_2$ symmetry is realized by identifying two sites.
Some models with the gauge symmetry
$G =$SU(2)$_{0}\times$SU(2)$_{1}\times  \cdots \times$SU(2)$_{2N}$ with identifying SU(2)$_{j}$ and SU(2)$_{2N-j}$ are equivalent to
the models in extra dimension on $S^1/Z_2$ upto $2N$ Kaluza-Klein (KK) modes.
The SU(2) sector in our model corresponds to the case for $N = 1$.
The similar approach was taken in studying a U(1) vector dark matter model~\cite{Abe:2012hb}.

Under this setup, we can write the Yukawa interaction terms as
\begin{align}
    - y_u \bar{q}_L \tilde{H} u_R
    - y_d \bar{q}_L H d_R   
    - y_e \bar{\ell}_L H e_R
    + (h.c.),
\label{eq:Yukawa}
\end{align}
where $\tilde{H} = \epsilon H^*$.
The gauge symmetry forbids $\Phi_1$ and $\Phi_2$ to couple to the fermions,
and only $H$ is the relevant Higgs field for the Yukawa interaction terms.
This Yukawa sector is as simple as one in the SM,
and we do not need to extend the fermion sector.
This is a reason why we add two extra SU(2) gauge symmetries into the SM.
If we added only one extra SU(2), there would be two possibilities.
One possibility is that the extra SU(2) is isolated and does not mix with the SU(2)$_L$ gauge field.
In this case, the dark SU(2) gauge bosons do not couple to the SM weak gauge bosons,
 and the model is the Higgs portal type. This is not our concern.
The other possibility is to mix the extra SU(2) gauge field with the SU(2) gauge field in the SM. It is expected by the mixing that the dark SU(2) gauge bosons couple to the SM weak gauge bosons. In this case, however, we need an exchanging symmetry under these two SU(2) gauge field to stabilize the dark matter. Since the SM left-handed fermions feel SU(2)$_L$ gauge symmetry, the symmetry exchanging the two SU(2) fields requires two types of the fermions; one is the doublet fields under an SU(2), the others are doublet under the other SU(2). Some linear combinations of them are the SM left-handed fermions, and the other linear combinations are extra fermions. Therefore, if we add only one extra SU(2), then the symmetry to stabilize the dark matter requires to double the fermion fields compared to the SM. 
On the other hand, by considering two extra SU(2) gauge symmetries, we can realize the simple Yukawa interaction terms without extending the fermion sector as in Eq.~\eqref{eq:Yukawa}. This is a distinctive feature of this model from other SU(2) dark matter models.

\subsection{Bosonic sector}
We briefly describe the electroweak sector and the related scalar sector.
More details are discussed in Appendices.
The Lagrangian for those two sectors is given by
\begin{align}
 {\cal L}\supset&
 - \frac{1}{4} B_{\mu \nu} B^{\mu \nu} 
 -\sum_{j=0}^2\sum_{a=1}^3 \frac{1}{4} W_{j\mu \nu}^a W_{j}^{a \mu \nu} 
 \nonumber\\
& + D_{\mu}H^\dagger D^{\mu} H 
  + \frac{1}{2} \tr{D_{\mu} \Phi_1^\dagger D^{\mu} \Phi_1}
  + \frac{1}{2} \tr{D_{\mu} \Phi_2^\dagger D^{\mu} \Phi_2}
 \nonumber\\
 &
 - V_{\text{scalar}},
\end{align}
where
\begin{align}
 V_{\text{scalar}}
=&
 m^2 H^\dagger H 
+ m_\Phi^2 \tr\left(\Phi_1^\dagger \Phi_1\right)
+ m_\Phi^2 \tr\left(\Phi_2^\dagger \Phi_2\right)
\nonumber\\
&
+ \lambda (H^\dagger H)^2
+ \lambda_\Phi \left(\tr\left(\Phi_1^\dagger \Phi_1\right)\right)^2
+ \lambda_\Phi \left(\tr\left(\Phi_2^\dagger \Phi_2\right)\right)^2
\nonumber\\
&
+ \lambda_{h\Phi} H^\dagger H  \tr\left(\Phi_1^\dagger \Phi_1\right)
+ \lambda_{h\Phi} H^\dagger H  \tr\left(\Phi_2^\dagger \Phi_2\right)
+ \lambda_{12} \tr\left(\Phi_1^\dagger \Phi_1\right) \tr\left(\Phi_2^\dagger \Phi_2\right).
\end{align}
Some coupling constants in the Higgs potential are common because of the discrete symmetry.
We assume that the Higgs fields obtain the following vacuum expectation values  at the global minimum.
\begin{align}
 \VEV{H}=& \begin{pmatrix} 0 \\ \frac{v}{\sqrt{2}} \end{pmatrix}, \quad
 \VEV{\Phi_1}= \VEV{\Phi_2}= \frac{1}{\sqrt{2}}\begin{pmatrix} v_\Phi & 0 \\ 0 & v_\Phi \end{pmatrix}.
\end{align}
The component fields of these Higgs fields at this vacuum are given by
\begin{align}
 H=& \begin{pmatrix} i\pi_3^+ \\ \frac{v + \sigma_3 - i \pi_3^0}{\sqrt{2}} \end{pmatrix},
\quad
 \Phi_j= 
   \begin{pmatrix} 
    \frac{v_\Phi + \sigma_j + i \pi_j^0}{\sqrt{2}} &  i \pi_j^+ \\
    i \pi_j^-   &  \frac{v_\Phi + \sigma_j - i \pi_j^0}{\sqrt{2}}   
    \end{pmatrix}.
\label{eq:component_fields}
\end{align}
From the stationary condition, we find
\begin{align}
m^2 =& -\lambda v^2 - 2 \lambda_{h\Phi} v_\Phi^2,\\ 
m_\Phi^2 =& -\frac{\lambda_{h\Phi}}{2} v^2 - (\lambda_{12}+ 2\lambda_\Phi) v_\Phi^2.
\end{align}

\subsection{Gauge sector}
After the electroweak symmetry breaking, the gauge boson mass terms are given by
\begin{align}
 \begin{pmatrix}
  W_{0\mu}^+ & W_{1\mu}^+ & W_{2\mu}^+
 \end{pmatrix}
 {\cal M}_{C}^2
 \begin{pmatrix}
  W_{0}^{-\mu} \\ W_{1}^{-\mu} \\ W_{2}^{-\mu}
 \end{pmatrix}
+
 \frac{1}{2}
 \begin{pmatrix}
  W_{0\mu}^3 & W_{1\mu}^3 & W_{2\mu}^3 & B_\mu
 \end{pmatrix}
 {\cal M}_{N}^2
 \begin{pmatrix}
  W_{0}^{3\mu} \\ W_{1}^{3\mu} \\ W_{2}^{3\mu} \\ B^\mu
 \end{pmatrix}
,
\end{align}
where
\begin{align}
  {\cal M}_{C}^2
=&
\frac{1}{4}
\begin{pmatrix}
 g_0^2 v_\Phi^2 & - g_0 g_1 v_\Phi^2 & 0 \\
-g_0 g_1 v_\Phi^2 &  g_1^2 (v^2 + 2 v_\Phi^2) & -g_1 g_0 v_\Phi^2 \\
 0  &  -g_1 g_0 v_\Phi^2 & g_0^2 v_\Phi^2
\end{pmatrix}
,\\
  {\cal M}_{N}^2
=&
\frac{1}{4}
\begin{pmatrix}
 g_0^2 v_\Phi^2 & - g_0 g_1 v_\Phi^2 & 0 & 0 \\
-g_0 g_1 v_\Phi^2 &  g_1^2 (v^2 + 2 v_\Phi^2) & -g_1 g_0 v_\Phi^2 & -g_1 g' v^2\\
 0  &  -g_1 g_0 v_\Phi^2 & g_0^2 v_\Phi^2 & 0 \\
 0 & -g_1 g' v^2 & 0 & g'^2 v^2
\end{pmatrix}
.
\end{align}
After diagonalizing these mass matrices, we find the following mass eigenstates,
\begin{align}
&\gamma, W^{\pm}, Z, \ V^0, V^{\pm}, \ W'^\pm, \ Z',
\end{align}
where $\gamma$, $W^{\pm}$, and $Z$ are identified as the SM electroweak gauge bosons.
$V^{0}$ and $V^{\pm}$ are odd under the discrete symmetry and are given by
\begin{align}
    V^{0}= \frac{W^{3}_{0\mu} - W^{3}_{2\mu}}{\sqrt{2}},
    \\
    V^{\pm}= \frac{W^{\pm}_{0\mu} - W^{\pm}_{2\mu}}{\sqrt{2}}.
\end{align}
The details, such as linear combinations for other gauge fields, are discussed in Appendix~\ref{app:gauge-sector}.

The masses of dark matter $V^{0}$ and its charged partner $V^{\pm}$ are given by
\begin{align}
 m_{V^\pm}^2 =  m_{V^0}^2 =& \frac{g_0^2 v_\Phi^2}{4} \equiv m_V^2,
\end{align}
at the tree level. At the loop level, the mass difference is generated, and $m_{V^\pm}$ becomes slightly heavier than $m_{V^0}$ as we discuss in Sec.~\ref{sec:mass-difference}.
Therefore, $V^0$ is a dark matter candidate in our model.

\subsection{Physical scalars}
There are 12 scalars in the model, and 9 of them are would-be NG bosons.
The three remaining neutral scalars are physical, and their mass terms are given by
\begin{align}
 {\cal L}\supset&
\frac{1}{2}
\begin{pmatrix}
 \sigma_3 & \sigma_1 & \sigma_2
\end{pmatrix}
\begin{pmatrix}
 2 \lambda v^2 & 2 v v_\Phi \lambda_{h\Phi} & 2 v v_\Phi \lambda_{h\Phi}\\
 2 v v_\Phi \lambda_{h\Phi} & 8 v_\Phi^2 \lambda_\Phi & 4 v_\Phi^2 \lambda_{12}\\
 2 v v_\Phi \lambda_{h\Phi} & 4 v_\Phi^2 \lambda_{12} & 8 v_\Phi^2 \lambda_\Phi
\end{pmatrix}
\begin{pmatrix}
 \sigma_3 \\ \sigma_1 \\ \sigma_2
\end{pmatrix}
.
\end{align}
After diagonalizing this mass matrix, we obtain the mass eigenstates, $h$, $h'$, and $h_D$, where $h_D$ is odd under the discrete symmetry.
\begin{align}
 \begin{pmatrix}
 \sigma_3 \\ \sigma_1 \\ \sigma_2 
\end{pmatrix}
=
\begin{pmatrix}
  \cos\phi_h & -\sin\phi_h & 0 \\
  \frac{1}{\sqrt{2}}\sin\phi_h  & \frac{1}{\sqrt{2}}\cos\phi_h & \frac{1}{\sqrt{2}} \\
 \frac{1}{\sqrt{2}}\sin\phi_h & \frac{1}{\sqrt{2}}\cos\phi_h     & -\frac{1}{\sqrt{2}}\\
\end{pmatrix}
\begin{pmatrix}
 h \\ h' \\ h_D
\end{pmatrix}
.
\end{align}
If we choose the mass eigenvalues and the mixing angle $(m_h, m_{h'}, m_{h_D}, \phi_h)$ as input parameters, then the quartic couplings in the Higgs potential are given by
\begin{align}
 \lambda=& \frac{m_h^2 \cos^2\phi_h + m_{h'}^2 \sin^2\phi_h}{2 v^2}, \label{eq:lambda}\\
 \lambda_{h\Phi}=& -\frac{\sin\phi_h \cos\phi_h}{2 \sqrt{2} v v_\Phi} (m_{h'}^2 - m_h^2),\\
 \lambda_{\Phi}=& \frac{m_h^2 \sin^2\phi_h + m_{h'}^2 \cos^2\phi_h + m_{h_D}^2}{16 v_\Phi^2},\\
 \lambda_{12}=& \frac{m_h^2 \sin^2\phi_h + m_{h'}^2 \cos^2\phi_h - m_{h_D}^2}{8 v_\Phi^2}.\label{eq:lambda12}
\end{align}

\subsection{Model parameters}
\label{eq:vphi_gg_v}
The Lagrangian in the electroweak sector contains the following parameters.
\begin{align}
\left(
g_0,\ g_1,\ g',\ m^2,\ m_\Phi^2,\ 
\lambda,\ \lambda_\Phi,\ \lambda_{h\Phi}, \lambda_{12}
\right).
\label{eq:parameters}
\end{align}
Instead of them, we can use the following parameters as inputs,
\begin{align}
\left(
e,\ m_Z,\ v,\ m_h,\ m_{Z'},\ m_V,\ 
 m_{h'},\ m_{h_D}, \phi_h
\right),
\label{eq:inputs}
\end{align}
where $e$ is the QED coupling constant, and $v$ is related to the Fermi constant as
\begin{align}
    v = \left( \sqrt{2} G_F \right)^{-1/2}.
    \label{eq:v-vs-Gf}
\end{align}
The first four parameters are already measured, and thus we have five free parameters in this model. The relation between the gauge couplings and the masses of the gauge bosons is discussed in Appendix~\ref{app:gauge-sector}.
The derivation of Eq.~\eqref{eq:v-vs-Gf} is discussed in Appendix~\ref{app:fermi-constant}.

The analytical expression of the relations between Eqs.~\eqref{eq:parameters} and \eqref{eq:inputs} is complicated.
In the following analysis, we numerically obtain the parameters in Eq.~\eqref{eq:parameters} from a given set of parameters in Eq.~\eqref{eq:inputs}. 
However, in some limits, these relations can be simplified.
Here we briefly show approximated expressions of some couplings for $v_\Phi \gg v$ that is typically realized for $m_{Z'} \gg m_{Z}$.
The approximate expressions help to understand the qualitative features of the model. 

We introduce $g_W$ as 
\begin{align}
 g_W
 \equiv
 \left( \frac{2}{g_0^2} + \frac{1}{g_1^2}\right)^{-1/2}.
\end{align}
We find $g_W \simeq 0.65 $ for $v_\Phi \gg v$ numerically,
namely $g_W$ is approximately the SU(2)$_L$ gauge coupling in the SM.
Using $g_W$, $m_{Z'}$, and $m_V$, we can obtain $g_0$, and $g_1$ as
\begin{align}
 g_0 \simeq& \sqrt{2} g_W \frac{m_{Z'}}{m_V} \frac{1}{\sqrt{\frac{m_{Z'}^2}{m_V^2}-1}}, \label{eq:g0}\\
 g_1 \simeq& g_W \frac{m_{Z'}}{m_V}. \label{eq:g1}
\end{align}
The mass ratio of $Z'$ and $V$ is given by
\begin{align}
 \frac{m_{Z'}^2}{m_V^2}
\simeq&
1 + \frac{2 g_1^2}{g_0^2}
.
\end{align}
This equation shows that $m_{Z'} > m_V$.
Using these approximations, 
we obtain the masses of $W$ and $W'$ as
\begin{align}
 m_W^2 \simeq& \frac{g_W^2 v^2}{4},\\
 m_{W'}^2 \simeq& m_{Z'}^2.
\end{align}

The gauge boson couplings to the fermions are given by 
\begin{align}
 g_{W u_L d_L}
= g_{W \ell_L \nu_L}(\equiv g_{W f_L f_L})
\simeq&
 g_W
,\\
 g_{W' u_L d_L}
=  g_{W' \ell_L \nu_L}(\equiv g_{W' f_L f_L})
\simeq&
 -g_W \sqrt{ \frac{m_{Z'}^2}{m_V^2} -1}
\label{eq:W'ff}
,\\
 g_{Z q_L q_L}
 = g_{Z \ell_L \ell_L}
 = g_{Z \nu_L \nu_L}
\simeq&
\frac{e}{s_Z c_Z} \left( t^3 -s_Z^2 Q \right)
,\\
 g_{Z' q_L q_L}
 = g_{Z' \nu_L \ell_L}
 = g_{Z' \nu_L \nu_L}
\simeq&
 - t^3  g_W \sqrt{ \frac{m_{Z'}^2}{m_V^2} -1}
\label{eq:Z'ff}
,\\
 g_{Z q_R q_R}
=  g_{Z \ell_R \ell_R}
=  g_{Z \nu_R \nu_R}
\simeq&
- \frac{e s_Z}{c_Z} Q 
,\\
 g_{Z' q_R q_R}
=  g_{Z' \ell_R \ell_R}
=  g_{Z' \nu_R \nu_R}
=&
{\cal O}\left( \frac{v^2}{v_\Phi^2}  \right)
,
\end{align}
where $t^3 = \dfrac{1}{2}$ $\left(-\dfrac{1}{2}\right)$ for up-type (down-type) fermions, $Q$ is the QED charge of the fermions, $c_Z = \sqrt{1-s_Z^2}$, and $s_Z$ is given as a solution of 
\begin{align}
    s_Z^2 c_Z^2 = \frac{v^2 e^2}{4 m_Z^2}.
\end{align}
We can see that the $W'$ and $Z'$ couplings to the SM fermions are controlled by 
the mass ratio of $Z'$ and $V$. 
If $m_{Z'}$ and $m_{V}$ are degenerated, then those couplings are suppressed
while $g_0$ becomes very large. 
Therefore, we expect that the values of $W'$ and $Z'$ couplings to the SM fermions are comparable to those of the $W$ couplings in the region where perturbation works.
We discuss this point further in Sec.~\ref{sec:constraint-on-mzp-mv}.

Using $g_W$ and the masses of the gauge bosons, 
we find that the triple gauge couplings are given by
\begin{align}
 g_{WWZ}
\simeq
 g_{W'W'Z} 
\simeq
 g_{V^-V^+Z} 
\simeq&
g_W \frac{m_W}{m_Z}
\simeq
g_{WWZ}^\text{SM}
\label{eq:V+V-Z}
,\\
 g_{WWZ'}
\simeq&
g_W \frac{m_W^2}{m_{Z'}^2}  \sqrt{\frac{m_{Z'}^2}{m_V^2} - 1}
,\\
 g_{WW'Z}
\simeq&
g_W \frac{m_W m_Z}{m_{W'}^2}  \sqrt{\frac{m_{Z'}^2}{m_V^2} - 1}
,\\
 g_{W'W'Z'} 
\simeq& 
g_W \frac{1}{\sqrt{\frac{m_{Z'}^2}{m_V^2} - 1}}
 \left( 2 - \frac{m_{Z'}^2}{m_V^2}\right)
,\\
 g_{WW'Z'}
=
 g_{W^+V^-V^0} =
 g_{W^-V^+V^0} 
\simeq& 
g_W
\label{eq:V0VpmWpm}
,\\
  g_{V^-W'^+V^0} 
= g_{W'^-V^+V^0} 
\simeq  g_{V^-V^+Z'} 
\simeq&
  g_W
\frac{1}{\sqrt{\frac{m_{Z'}^2}{m_V^2} - 1}}
.
\end{align}
We emphasize that 
$V^0$ and $V^\pm$ couple to $W$ and $Z$ without any  suppression factors, see Eqs.~\eqref{eq:V+V-Z} and \eqref{eq:V0VpmWpm}.
Therefore, 
DM pairs can annihilate into the SM gauge bosons through these couplings, $g_{V^-V^+ Z}$ and $g_{W^\pm V^\mp V^0}$.
This is a distinctive feature of our vector DM model.

Couplings of  physical scalar bosons to the gauge bosons are
\begin{align}
 g_{WWh}\simeq&
\frac{2 m_W^2}{v} \cos\phi_h
\simeq
g_{WWh}^\text{SM} \cos\phi_h
\label{eq:gWWh}
,\\
 g_{WW'h}\simeq&
\frac{2 m_W^2}{v} 
\left(
-\cos\phi_h \sqrt{\frac{m_{Z'}^2}{m_V^2} - 1}
+ \frac{m_W}{m_V} \frac{m_{Z'}}{m_V} \sin\phi_h
\right)
,\\
 g_{ZZh}\simeq&
\frac{2 m_Z^2}{v} \cos\phi_h
\simeq 
g_{ZZh}^\text{SM} \cos\phi_h
,\\
 g_{ZZ'h}\simeq&
\frac{2 m_W m_Z}{v}
\left(
- \cos\phi_h \sqrt{ \frac{m_{Z'}^2}{m_V^2} - 1}
+ \frac{m_W}{m_V} \frac{m_{Z'}}{m_V} \sin\phi_h
\right)
,\\
 g_{V^0 V^0h}\simeq  g_{V^+ V^- h}\simeq&
\frac{g_W m_{Z'}}{\sqrt{\frac{m_{Z'}^2}{m_V^2}-1}} \sin \phi_h
,\\
 g_{V^0 V^0h'}\simeq  g_{V^+ V^-h'}\simeq&
\frac{g_W m_{Z'}}{\sqrt{\frac{m_{Z'}^2}{m_V^2}-1}} \cos \phi_h
.
\end{align}
Note that $g_{WWh}$ is the same as the SM prediction for $\cos \phi_h = 1$.
This $g_{WWh}$ coupling is already measured by the ATLAS and CMS experiments,
and the measured value is consistent with the SM value. 
Accordingly, we take small $\phi_h$ in the following analysis.
For a small $\phi_h$ limit, the $V^0$ coupling to $h$ is suppressed. 
However, as we mentioned already, the annihilation processes of DM pairs into the SM particles do not need to rely on the DM-Higgs coupling. 
Therefore, we can obtain the right amount of DM energy density however small $|\phi_h|$ we take.

\section{Constraints}\label{sec:constraints}
\subsection{Perturbative unitarity}

We obtain the constraints on $g_0$, $g_1$, and scalar quartic couplings from the perturbative unitarity conditions for two-particle scattering processes in the high energy regime.

First, we consider two-to-two scalar bosons scattering processes in the high energy limit and derive the constraints on the scalar quartic couplings.
In our derivation, we assume that these quartic couplings are much larger than the other couplings, such as gauge couplings.
This model contains $12$ scalars 
and there are $76$ two scalar particle channels. 
We obtain the following conditions. 
\begin{align}
&|\lambda| \leq 4\pi,\\
&|\lambda_{h\Phi}| \leq 4\pi,\\
&|\lambda_{\Phi}| \leq \pi,\\
&|\lambda_{12}| \leq 2\pi,\\
&|3\lambda_{\Phi}-\lambda_{12}| \leq \pi,\\
&\left|3\lambda + 4(3\lambda_\Phi + \lambda_{12}) \pm \sqrt{\left(3\lambda - 4(3\lambda_\Phi + \lambda_{12}) \right)^2 + 32 \lambda_{h\Phi}^2}\right| \leq 8\pi.
\end{align}

Second, we can derive the upper bounds on the gauge couplings from vector-vector to scalar-scalar scattering processes. In our model, one of $g_0$ and $g_1$ can be larger than the other in most of the region of the parameter space, and thus the result in Ref.~\cite{1202.5073} is applicable. We find that
\begin{align}
g_j < \sqrt{\frac{16 \pi}{\sqrt{6}}} 
\simeq 4.53.
\quad \left(j = 0, 1\right)
\label{eq:PU-for_g_j}
\end{align}

\subsection{The mass ratio of $Z'$ and $V$}
\label{sec:constraint-on-mzp-mv}
We find in Sec.~\ref{eq:vphi_gg_v} that
the mass ratio of $Z'$ and $V$ is important to determine 
the model parameters and couplings.
Although the mass ratio is a free parameter, there is a viable range.

It can be seen from Eq.~\eqref{eq:g0} that
$g_0$ becomes very large for $m_{Z'} \sim m_V$, and we can not treat $g_0$ as a small perturbation.
For $m_{Z'} \gg m_V$, 
we can see from Eqs.~\eqref{eq:g1} and \eqref{eq:W'ff} 
that $g_1$ and $g_{W'f_L f_L}$ become large.
This is also bad for the perturbative calculation.
Moreover, the decay width of $W'$ and $Z'$ becomes larger for the larger $g_{W'f_L f_L}$. 

For $v_\Phi \gg v$ and $|\phi|\ll 1$, we find 
\begin{align}
 \Gamma(W' \to f\bar{f}) 
\simeq &
\frac{N_c}{48\pi} m_{W'} g_{W f_L f_L}^2 
\left(  \frac{m_{Z'}^2}{m_V^2} -1 \right)
\label{eq:W'2ff}
,\\
\Gamma(W' \to WZ)
\simeq&
 \frac{1}{192\pi} m_{W'}  g_{W f_L f_L}^2
\left(  \frac{m_{Z'}^2}{m_V^2} -1 \right)
\label{eq:W'2WZ}
,\\
\Gamma(W' \to Wh)
\simeq&
 \frac{1}{192\pi} m_{W'} 
g_{W f_L f_L}^2
\left(  \frac{m_{Z'}^2}{m_V^2} -1 \right)
\label{eq:W'2Wh}
,
\end{align}
where $N_c = 3$ for quarks and 1 for leptons.
Here we take $V_{CKM} = 1$ for simplicity. 
If $W'$ cannot decay into the non-SM particles kinematically,
then the total width of $W'$ is given by 
\begin{align}
 \Gamma_{W'}
\simeq&
m_{W'}
\frac{25}{96\pi}
g_{W}^2
\left(\frac{m_{Z'}^2}{m_V^2} - 1\right)
.
\label{eq:GammaW}
\end{align}
We show some values of $g_0$, $g_1$, $\Gamma_{W'}/m_{W'}$, and $|g_{W'f_L f_L}/g_{Wf_L f_L}|$ for given ratios of $m_{Z'}$ and $m_V$ in Tab.~\ref{tab:Wp-total-width__gWpff}.
We find that we cannot treat $g_0$ as a small perturbation for $m_{Z'} \simeq m_V$.
We obtain a lower bound on the ratio of masses of $Z'$ and $V$ as $m_{Z'}/m_V \gtrsim 1.02$ from the perturbativity condition for $g_0$ shown in Eq.~\eqref{eq:PU-for_g_j}. 
Similarly, the perturbativity for $g_1$ gives an upper bound on $m_{Z'}/m_V$.
We find $m_{Z'}/m_V < 6.97$.
The total width also gives an upper bound on $m_{Z'}/m_V$
because the total width is proportional to the imaginary part of the one-loop diagrams while the mass is at the tree level.
Therefore, our calculation based on the perturbation is valid only for the region where $m_{W'} > \Gamma_{W'}$.
This gives the upper bound on $m_{Z'}$ for a given value of $m_V$, and we find that $m_{Z'} < 5.45 m_V$.
We also find that $\Gamma_{W'}/m_{W'} < 0.1$ is satisfied for $m_{W'} \lesssim 2 m_V$.
\begin{table}[tb]
\centering
\caption{
The values of $g_0$, $g_1$, $\Gamma_{W'}/m_{W'}$, and $g_{W'f_L f_L}/g_{Wf_L f_L}$ given ratios of $m_{Z'}$ and $m_V$.}
\label{tab:Wp-total-width__gWpff}
\setlength\tabcolsep{8.5pt} 
 \begin{tabular}{ccccc}\hline
 $m_{Z'}/m_V$ &  $g_0$  & $g_1$ & $\Gamma_{W'}/m_{W'}$ & $|g_{W'f_L f_L}/g_{W f_L f_L}|$ \\ \hline \hline
 1.02 & 4.53 & 0.661 & 0.00148 & 0.207 \\
 1.05 & 3 & 0.680 & 0.00358 & 0.321 \\
 $\sqrt{2}$ & 1.30 & 0.916 & 0.0348 & 1 \\
 4.63 & 0.938 & 3 & 0.711 & 4.52 \\
 5.45 & 0.932 & 3.53 & 1 & 5.36 \\ 
6.97 & 0.93 & 4.53 & 1.66 & 6.90 \\ \hline
 \end{tabular}
\end{table}

\subsection{$W'$ and $Z'$ searches at the LHC}

New heavy vector bosons are being searched by the ATLAS and CMS experiments.
Our model predicts the heavy vector bosons, $W'$ and $Z'$, 
and they couple to the SM particles.
The $W'$ and $Z'$ couplings to SM particles are determined by the ratio of $m_{Z'}$ and $m_V$ as discussed in Sec.~\ref{eq:vphi_gg_v}.
The couplings to the fermions and the SM vector bosons can be as large as the SU(2)$_L$ gauge coupling in the SM,
and the former is larger than the latter.
Therefore, the main production process of $W'$ and $Z'$ at the LHC is $q\bar{q} \to W'/Z'$. The branching fraction to two fermions is larger than two bosons, see Eqs.\eqref{eq:W'2ff}--\eqref{eq:W'2Wh}.
Therefore, the main search channel of $W'$ and $Z'$ are 
$pp \to W' \to \ell \nu$ and $pp \to Z' \to \ell \ell$.
The former gives the stronger constraint on the mass of $W'$, and we focus on that process here. 

\begin{figure}[tb]
\centering
\includegraphics[width=0.68\hsize]{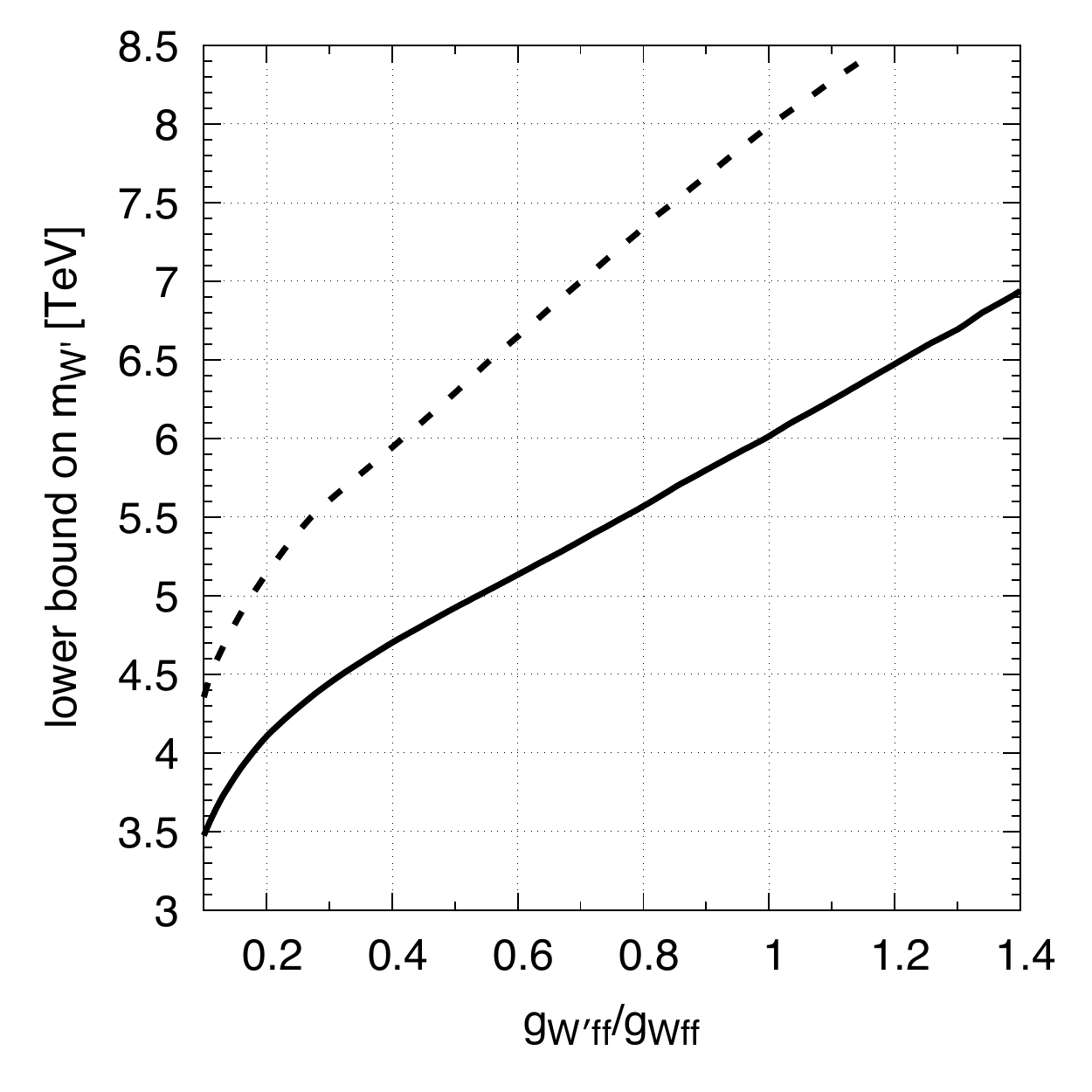}
\caption{The solid curve shows the lower bound on $m_{W'}$ for a given $g_{W'f_L f_L}$ coupling obtained by recasting the result in Ref.~\cite{1906.05609}.
The dashed curve shows the prospect at the ATLAS experiment with $14$~TeV with $3000$~fb$^{-1}$\cite{ATL-PHYS-PUB-2018-044}.
}
\label{fig:W'recast}
\end{figure}

The ATLAS experiment searches the $pp \to W' \to \ell \nu$ process and
finds the lower bound on $m_{W'}$ as 6~TeV for the Sequential Standard Model (SSM)~\cite{1906.05609}.\footnote{The CMS experiment also searches the same channel but gives a weaker bound on $m_{W'}$, $m_{W'} > 5.2$~TeV~\cite{Sirunyan:2018mpc}.}
The $W'$ couplings to the SM fermions in our model are different from those in the SSM.
We recast the bound and obtain the lower bound on $m_{W'}$ for a given coupling 
ratio of $g_{W' f_L f_L}$ and $g_{W f_L f_L}$. 
The result is shown in Fig.~\ref{fig:W'recast}.
Here we assume that the $K$ factor is 1.3.
We find that
$m_{W'} \gtrsim 7$~TeV for $g_{W' f_L f_L}/g_{W f_L f_L} \gtrsim 1.42$.
Since the ATLAS experiment does not give the bound for $m_{W'} > 7$~TeV,
we cannot obtain the bound on $m_{W'}$ for $g_{W' f_L f_L}/g_{W f_L f_L} \gtrsim 1.42$.
Similarly, we also recast the prospect of $W'$ search at the ATLAS experiment with 14~TeV with 3000~fb$^{-1}$~\cite{ATL-PHYS-PUB-2018-044}.
Other channels give weaker bound than this $\ell \nu$ channel.

\subsection{Electroweak precision measurements}

For $m_{W'/Z'} \gg m_{W/Z}$ limit, it is easy to obtain the electroweak precision 
parameters, $\hat{S}$, $\hat{T}$, $W$, and $Y$, introduced in~\cite{hep-ph/0405040}.
At the tree level, we find that
\begin{align}
 &\hat{S} = \hat{T} = Y = 0,\nonumber\\
 & W = \frac{2 g_1^2}{g_0^2 + 2 g_1^2} \frac{m_W^2}{m_{W'}^2}.
\simeq 
\left(1 - \frac{m_V^2}{m_{Z'}^2}\right) \frac{m_W^2}{m_{W'}^2}
.
\end{align}
The constraint is given as
$ W = (-0.3 \pm 0.6) \times 10^{-3}$.
We find that this constraint is much weaker than the constraint from the $W'$ search at the LHC experiment.

\subsection{Higgs signal strength}
\label{sec:HiggsSignalStrength}
Among the three scalar fields, only $H$ contributes to the Yukawa interaction terms,
and thus the $h$ couplings to the fermions are equal to those in the SM times $\cos\phi_h$. 
As we have shown in Eq.~\eqref{eq:gWWh},
$g_{WWh}$ for $v_\Phi \gg v$ is approximately given by the SM coupling times $\cos\phi_h$. 
Thus the Higgs signal strengths are given by
\begin{align}
\kappa_F = \cos\phi_h,
\quad
\kappa_V \simeq \cos\phi_h.
\end{align}
We can constrain $\phi_h$ from the measurement of the Higgs couplings.
We use the result from the ATLAS experiment~\cite{1909.02845}, 
\begin{align}
    \kappa_V =& 1.05 \pm 0.04,\\
    \kappa_F =& 1.05 \pm 0.09,
\end{align}
with the linear correlation between them is observed as 44\%,
and obtain $|\phi_h| < 0.3$. 
We consider $0 \leq |\phi_h| < 0.3$ in the following discussions.

\section{DM phenomenology}
\label{sec:DM-pheno}

\subsection{Mass difference and its implication for collider physics}
\label{sec:mass-difference}
At the tree level, $V^0$ and $V^\pm$ have the same mass.
However, the mass difference is generated at the loop level, 
and thus $V^\pm$ is slightly heavier than $V^0$.
The mass difference is given by
\begin{align}
\delta_{m_V}
\equiv
 m_{V^\pm} - m_{V^0}
=&
  \sqrt{m_V^2 + \Pi_{V^+V^-}(m_{V^\pm}^2)}
- \sqrt{m_V^2 + \Pi_{V^0V^0}(m_{V^0}^2)}
\nonumber\\
\simeq&
  \frac{\Pi_{V^+V^-}(m_V^2)- \Pi_{V^0V^0}(m_V^2)}{2 m_V} 
,
\end{align}
where $\Pi_{V^+V^-}$ and $\Pi_{V^0V^0}$ are the self-energies of $V^\pm$ and $V^0$, 
respectively.
We calculate $\delta_{m_V}$ at the one-loop level by using
\texttt{FormCalc} \cite{hep-ph/9807565}.
In $v_\Phi \gg v$ limit, we find
\begin{align}
 \delta_{m_V}
\simeq
\frac{m_W^3 G_F}{\sqrt{2}\pi}
\left( 1 - \frac{m_W}{m_Z} \right)
\simeq 
168~\text{MeV}
.
\end{align}
This result is consistent with the result in~\cite{hep-ph/0512090}.
We have also checked it numerically by using \texttt{LoopTools}~\cite{hep-ph/9807565},
without taking $v_\Phi \gg v$ limit.

\begin{figure}[tbp]
\begin{minipage}{1\hsize}
   \begin{minipage}[t]{0.49\hsize}
       \begin{center}
       \includegraphics[width=7cm,pagebox=cropbox,clip]{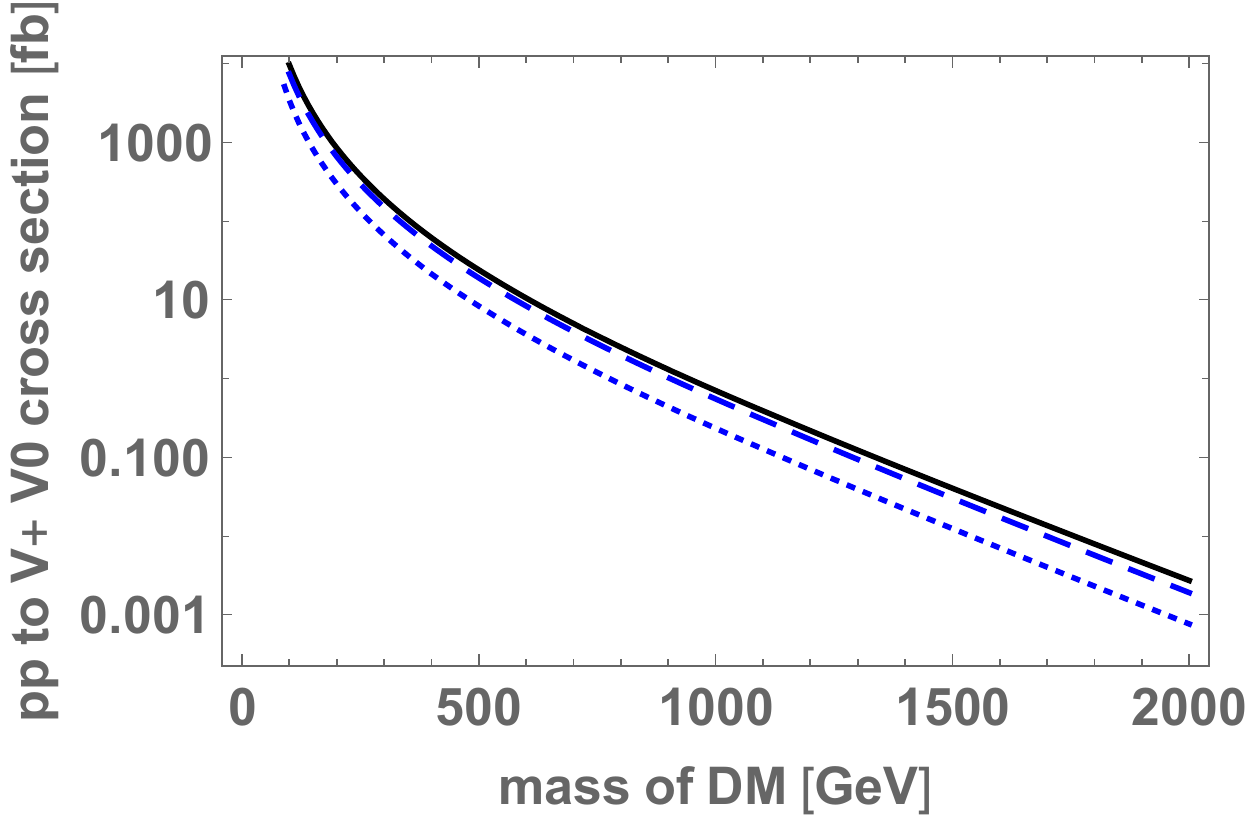}
       \end{center}\label{VpV0}
   \end{minipage}
   \begin{minipage}[t]{0.49\hsize}
       \begin{center}
       \includegraphics[width=7cm,pagebox=cropbox,clip]{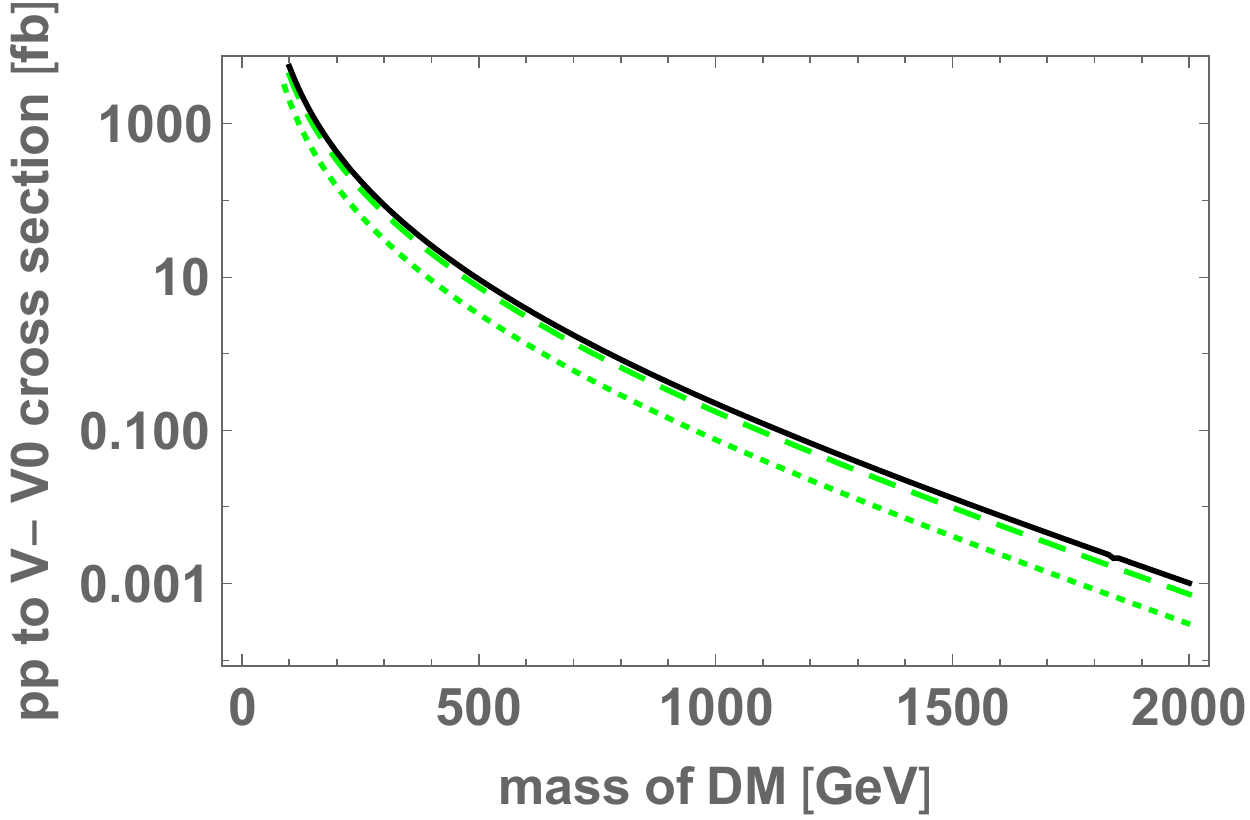}
       \end{center}\label{VmV0}
    \end{minipage}
       \begin{center}
       \includegraphics[width=7cm,pagebox=cropbox,clip]{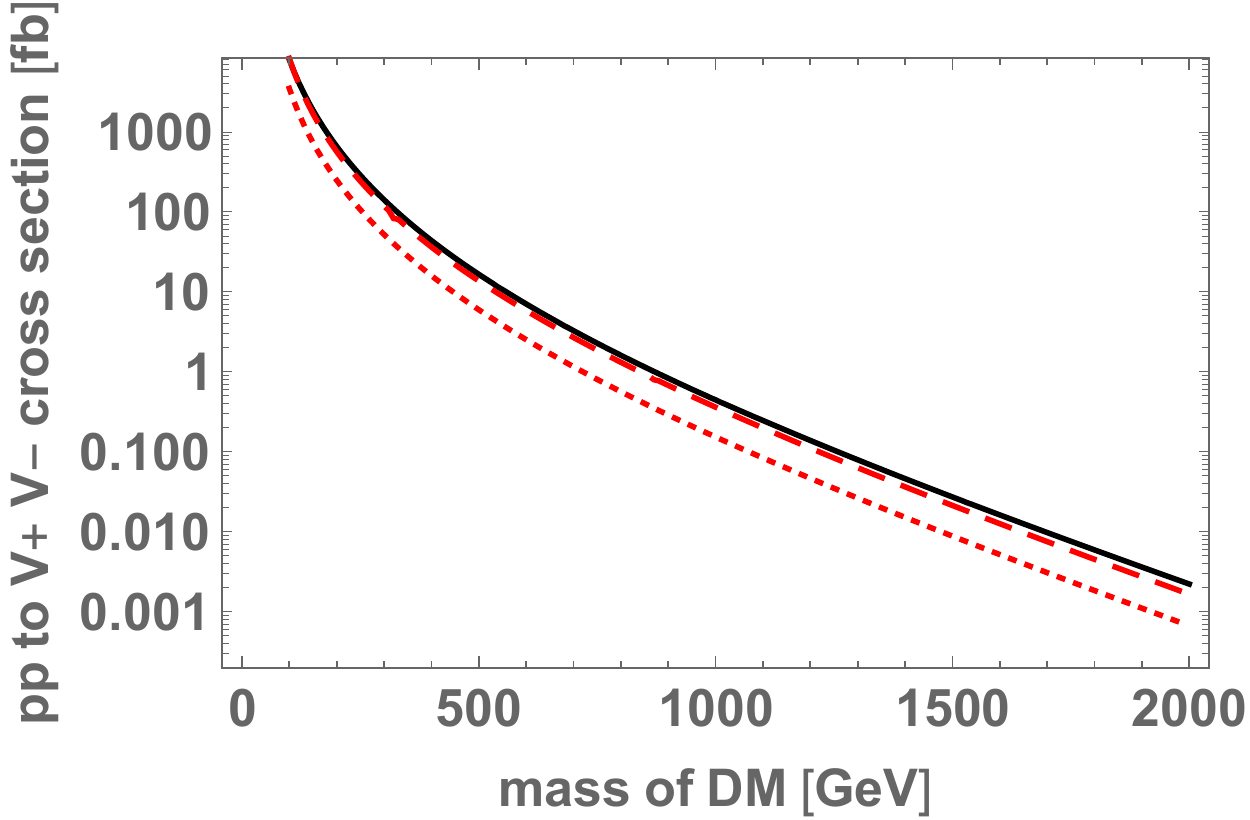}
       \end{center}\label{VpVm}
\end{minipage}
\caption{The production cross section of $V^\pm$ and $\tilde{W}^\pm$ from proton collisions at $\sqrt{s} = 13$~TeV. 
The left panel shows the cross section of $pp \to V^+ V^0$, 
the right shows $pp \to V^- V^0$, 
and the bottom shows $pp \to V^+ V^-$.
In each figure, the black line shows the Wino production cross section and the dashed (dotted) line shows the $V^\pm$ for $m_{Z',W'}=1.5 m_V$ ($m_{Z',W'}=1.3 m_V$).}\label{VVpro}
\end{figure}

This small mass difference is the same as the mass difference between the charged and neutral components of Wino $(\tilde{W})$ in the MSSM. Wino is SU(2)$_L$ triplet fermions. The charged Wino decays into the neutral Wino, but its lifetime is long due to the small mass difference.
Thus, Wino is being searched in the long-lived particle searches at the LHC. 
Our DM candidate, $V^0$, and its partner, $V^\pm$, has the same properties as the Wino. 
The decay rate of $V^\pm$ and the mass difference of $V^\pm$ and $V^0$ are exactly equal to those of Wino. Therefore, the long-lived particle search is also a useful tool to find $V^\pm$ in our model.
The only difference of $V$ from $\tilde{W}$ is the production rate of the charged particles.
Figure~\ref{VVpro} shows the production cross sections of $V^{\pm}$ and $\tilde{W}^{\pm}$ at the LHC with $\sqrt{s} = 13$~TeV.
We find that the production cross section of $V^{\pm,0}$ depends on $m_{W'}$ and $m_{Z'}$ as well as $m_V$.
It is also found that
the production cross section of $V^{\pm,0}$ is smaller than the production cross section of Wino
because of the interference between the diagrams exchanging $W$ and $W'$ ($Z$ and $Z'$) in the $s$-channel.
Therefore, the constraint on $m_V$ from the long-lived particle search is weaker than that on the Wino, 
$m_{\tilde{W}} \gtrsim 460 $~GeV~\cite{1712.02118}.
Once we require $V^0$ to explain the measured value of the DM energy density, then $m_V \gtrsim$ 3~TeV is required as we will see in the following. Therefore, our model is consistent with the results of the long-lived search if the whole of DM in our universe is explained by $V^0$.

\subsection{Direct detection}

At the leading order, DM-nucleon scattering is mediated by two scalars, $h$ and $h'$, which are even under the discrete symmetry.
The spin-independent vector DM-nucleon scattering cross section is given by
\begin{align}
\sigma_{\rm SI}^N = \frac1\pi  \left( \frac{m_N}{m_N + m_{V}} \right)^2 |f_{NV}|^2 ,
\end{align}
where $m_N$ is the nucleon mass ($N=p, n$) and $f_{NV}$ is the effective coupling of DM-nucleon interactions.

\begin{figure}[tbp]
\centering
\begin{minipage}{1\hsize}
   \begin{minipage}[t]{0.49\hsize}
       \begin{center}
       \includegraphics[height=3.5cm,pagebox=cropbox,clip]{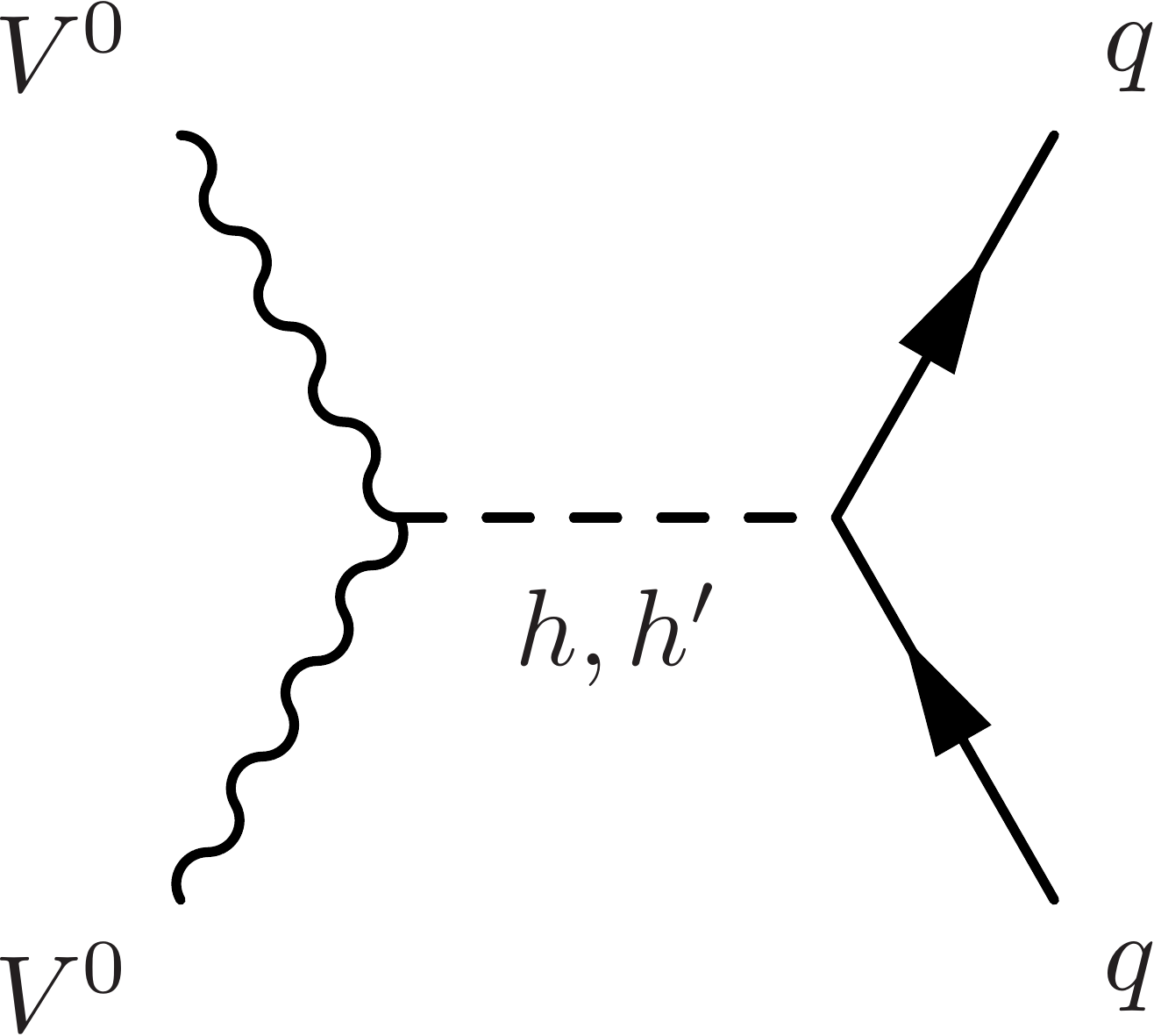}
       \end{center}
   \end{minipage}
   \begin{minipage}[t]{0.49\hsize}
       \begin{center}
       \includegraphics[height=3.5cm,pagebox=cropbox,clip]{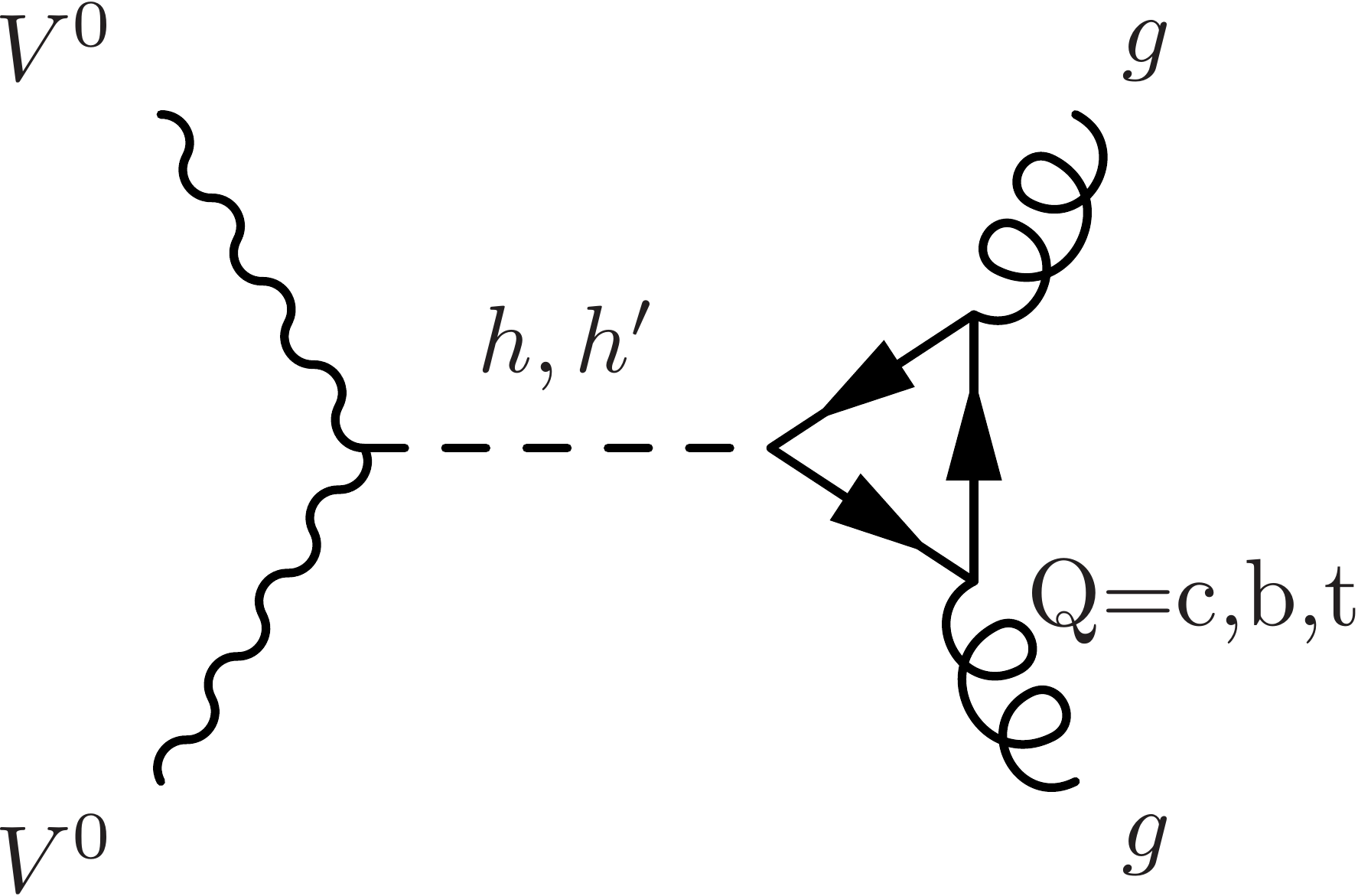}
       \end{center}
    \end{minipage}
\caption{The leading diagrams mediated by $h$ and $h'$.  }\label{DDdiagram}
\end{minipage}
\end{figure}

Figure~\ref{DDdiagram} shows the leading diagrams at the parton-level.
The following Parton-level effective interactions are relevant to the DM-nucleon cross section,
\begin{align}
\mathcal{L}^{\rm eff} = 
& \sum_{q=u,d,s} c_q V^{0 \mu} V^0_\mu m_q \bar{q} q 
    + \sum_{Q=c,b,t} c_Q V^{0\mu} V_\mu^0 m_Q \bar{Q} Q,
\end{align}
where $m_q$ and $m_Q$ are light and heavy quark masses, respectively.
The couplings, $c_q$ and $c_Q$, in our model are
\begin{align}
c_q = c_Q 
&= \frac{m_{V}^2}{\sqrt{2} v v_\Phi} \sin\phi_h \cos\phi_h \left( \frac1{m_{h'}^2} - \frac1{m_h^2} \right).
\end{align}

To obtain the effective coupling of the DM-nucleon interactions, $f_{NV}$, we use the nucleon matrix elements,
\begin{align}
\langle N | m_q \bar{q} q | N \rangle &\equiv m_N f^{(N)}_{Tq}, ~~~(q=u,d,s),\\
\langle N | \frac{\alpha_s}{\pi} G^a_{\mu\nu} G^{a \mu\nu} |
N \rangle &= - \frac{8}{9} m_N \left( 1 - \sum_q f^{(N)}_{Tq} \right).
\end{align}
where $G^a_{\mu\nu}$ and $\alpha_s$ are the SU(3)$_c$ field strength tensor and coupling constant, respectively.  
The numerical values of the mass fractions for the nucleon,
$f^{(N)}_{Tq} (N=p,n)$, are obtained by lattice simulations, 
and we take the default values of \texttt{micrOMEGAs} \cite{1801.03509}.
\begin{align}
\begin{aligned} \label{matrixlat}
f^p_{Tu} &= 0.0153,\  f^n_{Tu} = 0.011 \\
f^p_{Td} &= 0.0191,\  f^n_{Td} = 0.0273 \\
f^{p}_{Ts} &= f^{n}_{Ts} = 0.0447.
\end{aligned} 
\end{align}
For light quarks ($q=u,d,s$), we can obtain the contribution to the effective coupling $f_{NV}$ using nucleon matrix elements of the mass operators.
For the heavy quarks ($Q = c,b,t$), the leading contribution is loop diagrams (Fig.~\ref{DDdiagram} right).
The operator $m_Q \bar{Q} Q$ equals $- \frac{\alpha_s}{12\pi} G^a_{\mu\nu} G^{a\mu\nu} $ in the matrix element, so the matrix elements of the heavy quark mass operators are given by
\begin{align}
\langle N | m_Q \bar{Q} Q | N \rangle = \frac2{27} m_N \left(1 - \sum_q f^{(N)}_{Tq} \right), ~~~(Q=c,b,t).
\end{align}
Using these matrix elements, the effective coupling $f_{NV}$ is given by
\begin{align}
\frac{f_{NV}}{m_N} 
=& \sum_q c_q f_{Tq}^N + \frac2{27} \sum_Q c_Q (1 - \sum_q f_{Tq}^N) \nonumber
\\
=& \frac{m_{V}^2}{\sqrt{2} v v_\Phi} \sin\phi_h \cos\phi_h \left( \frac1{m_{h'}^2} - \frac1{m_h^2} \right) \left( \frac29 + \frac79 \sum_q f^{(N)}_{Tq} \right).
\end{align}
Finally, we obtain the spin-independent nucleon-vector DM cross section as follows.
\begin{align}
\sigma_{\rm SI}^N =& 
\frac{1}{2 \pi} \frac{m_N^4 m_{V}^4}{(m_N + m_{V})^2} \frac1{v^2 v_\Phi^2} \sin^2{\phi_h} \cos^2{\phi_h} \left( \frac1{m_{h'}^2} - \frac1{m_h^2} \right)^2 \left( \frac29 + \frac79 \sum_{q=u,d,s} f^N_{Tq} \right)^2
\nonumber
\\
\simeq& 
\frac{g_0^2}{32 \pi v^2} \frac{m_N^4}{m_h^4} 
\sin^2(2 \phi_h)
\left( \frac29 + \frac79 \sum_{q=u,d,s} f^N_{Tq} \right)^2 
\nonumber
\\
\simeq& 
10^{-44} \times
g_0^2 \sin^2(2 \phi_h) 
\text{\ [cm}^2\text{]}.
\label{SIcs}
\end{align}
Here we assumed that $m_V \gg m_N$, and also $m_{h'} \gg m_h$ in the last two lines of Eq.~(\ref{SIcs}).
This cross section is proportional to $\sin^2 (2\phi_h)$, and thus the large $|\phi_h|$ region is severely constrained from the direct detection experiments.
The direct detection limit on the DM-nucleon cross section for TeV scale DM is around $10^{-45}$~cm$^2$~\cite{1805.12562}.
For $g_0 = 1$, we find $\phi_h \lesssim 0.15$.
This upper bound can be stronger than the bound from the Higgs signal strength.
If $\phi_h$ is smaller than $\sim 0.01$, the higher-order diagrams dominate in the DM-nucleon SI scattering process  so  that $\sigma_{\rm SI}^N\sim 10^{-47}$cm$^2$ \cite{Hisano:2004pv, Hisano:2010fy, Hisano:2015rsa} .

\subsection{Relic abundance}

The model contains two DM candidates, $V^0$ and $h_D$.
In this paper, we treat $V^0$ as the DM candidate by assuming $h_D$ is always heavier than $V^0$.

We calculate the thermal relic abundance of $V^0$ by using
\texttt{micrOMEGAs}~\cite{1801.03509}.
The model file is generated by \texttt{FeynRules}~\cite{1310.1921}.
Since the mass difference of $V^{\pm}$ and $V^0$ is tiny,
the coannihilation processes, which are automatically calculated in
\texttt{micrOMEGAs}, are relevant.
All the masses of the new particles are proportional to $v_\Phi$,
hence the large mass difference among the new particles requires large couplings.
To avoid large couplings and to keep working within the perturbative regime,
we keep the mass ratio of the new particles to the DM mass within ${\cal O}(1)$.

The vector DM can interact with the SM weak gauge bosons even in a limit of vanishing the scalar mixing $\phi_h$. 
We start by investigating the relic abundance with very small $\phi_h$ and show that the vector DM can explain the measured value of the DM energy density. 
We also discuss the case for $|\phi_h| \sim {\cal O}(0.1)$ to see the impact of $\phi_h$ on the forthcoming direct detection experiments.

\subsubsection{Very small $|\phi_h|$ case}

\begin{figure}[tb]
\centering
\includegraphics[width=0.68\hsize]{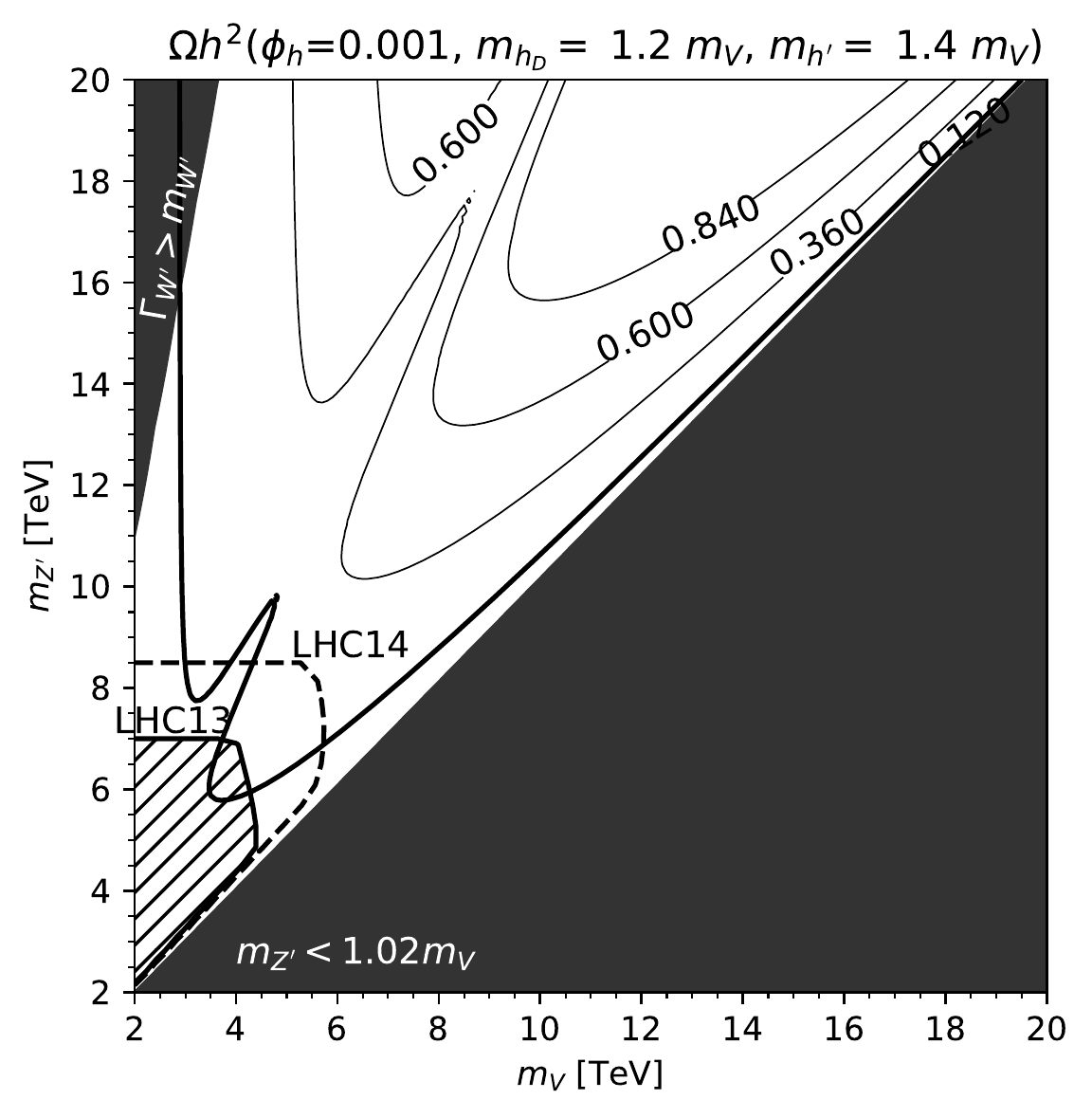}
\caption{The contours show the DM relic abundance as a function of the masses of the DM and $Z'$. Here, 
$\phi_h =0.001$, $m_{h_D} = 1.2~m_V$ and $m_{h'} = 1.4~m_V$.
The measured value of the DM energy density is shown by the thick-solid contour. 
The DM is overabundant in the region above the thick-solid contour. 
The region filled by the hatched pattern is excluded by the ATLAS experiment~\cite{1906.05609}.
The prospect at the HL-LHC is also shown by the dashed curve~\cite{ATL-PHYS-PUB-2018-044}.
In the larger black-shaded region, $g_0$ is beyond the perturbative unitarity bound, see Tab.~\ref{tab:Wp-total-width__gWpff}.
In the black-shaded region in the left-top corner, $\Gamma_{W'} > m_{W'}$.
}
\label{fig:abundance}
\end{figure}

Figure~\ref{fig:abundance} shows the DM relic abundance
in an $m_V$-$m_{Z'}$ plane for the very small $|\phi_h|$.
We take $\phi_h =0.001$ here, and the same result is obtained for much smaller $\phi_h$.
This is because the hidden vector bosons, $V^0$ and $V^\pm$,  efficiently annihilate into visible vector bosons and do not need to rely on $h$ and $h'$ exchanging processes.
The other new particle masses are fixed as $m_{h_D} = 1.2~m_V$ and $m_{h'} = 1.4~m_V$.
The result is insensitive to the choice of $m_{h_D}$ and $m_{h'}$, $m_{h_D} = 1.2~m_V$ and $m_{h'} = 1.4~m_V$.
We find three viable regions of parameter space for the explanation of the measured value of the DM energy density~\cite{1807.06209} as a thermal relic:
the narrow $W'$ width region ($m_{Z'} \lesssim 2~ m_V$), 
the $V'$-resonant region ($m_{Z'} \simeq 2~m_V$), 
and the wide $W'$ width region ($m_{Z'} \gtrsim 2~m_V$). 

For the narrow $W'$ width region, pairs of the dark vector bosons mainly annihilate into visible massive gauge bosons including $W'^\pm$ and $Z'$.
In this region, we find $m_{V} \gtrsim 4.2$~TeV from
the constraint on the $W'$ search by the ATLAS experiment~\cite{1906.05609}.
It is possible to test this case for $m_{V} \lesssim 5.8$~TeV by the $W'$ search at the HL-LHC~\cite{ATL-PHYS-PUB-2018-044}.
For the larger $m_V$, we can avoid the constraint from the $W'$ and $Z'$ search because it requires heavier $W'$ and $Z'$ to obtain the measured value of the DM energy density. However, it also requires the larger $g_0$, and thus the perturbative unitarity of $g_0$ gives the upper bound on $m_V$.

In the $V'$-resonant region, which looks like a horn in the figure, the main (co)annihilation channel is
$V^0 V^\pm \to q \bar{q}$ via $W'$ exchange in the $s$-channel. 
In this region, $g_0$ and $g_1$ are less than ${\cal O}(1)$,
and the perturbative unitarity is easily satisfied.

In the wide $W'$ width region, pairs of the dark matter particles mainly annihilate into $W^\pm$ and $Z$ because the processes with a $W'^\pm$ or a $Z'$ in final states are kinematically forbidden in this region. The masses of $W'$ and $Z'$ are larger than the dark matter particles, and thus $W'$ and $Z'$ are almost decoupled from the annihilation processes. As a result, $m_V$ is almost fixed around 3~TeV if we demand $\Omega h^2 = 0.12$.
This region is similar to the Wino DM model and SU(2)$_L$ triplet scalar DM models~\cite{Hisano:2006nn,Cirelli:2005uq}. In those models, DM mainly annihilates into $W^\pm$ and $Z$, and the mass of the DM is fixed by requiring the thermal relic to explain the measured value of the DM energy density.

\subsubsection{For $|\phi_h| \sim {\cal O}(0.1)$ }
We discuss the case for $|\phi_h| \simeq {\cal O}(0.1)$ to see the effects of $\phi_h$ to the thermal relic abundance and the direct detection experiments.
In this regime,
the scalar quartic couplings can be large with large $m_{h'}$
as can be seen from Eqs.~\eqref{eq:lambda}--\eqref{eq:lambda12}.
The annihilation processes into $h$ and $h'$, which are proportional to the quartic couplings, are efficient, and $\phi_h$ dependence is visible. 

\begin{figure}[tb]
\centering
\includegraphics[width=0.68\hsize]{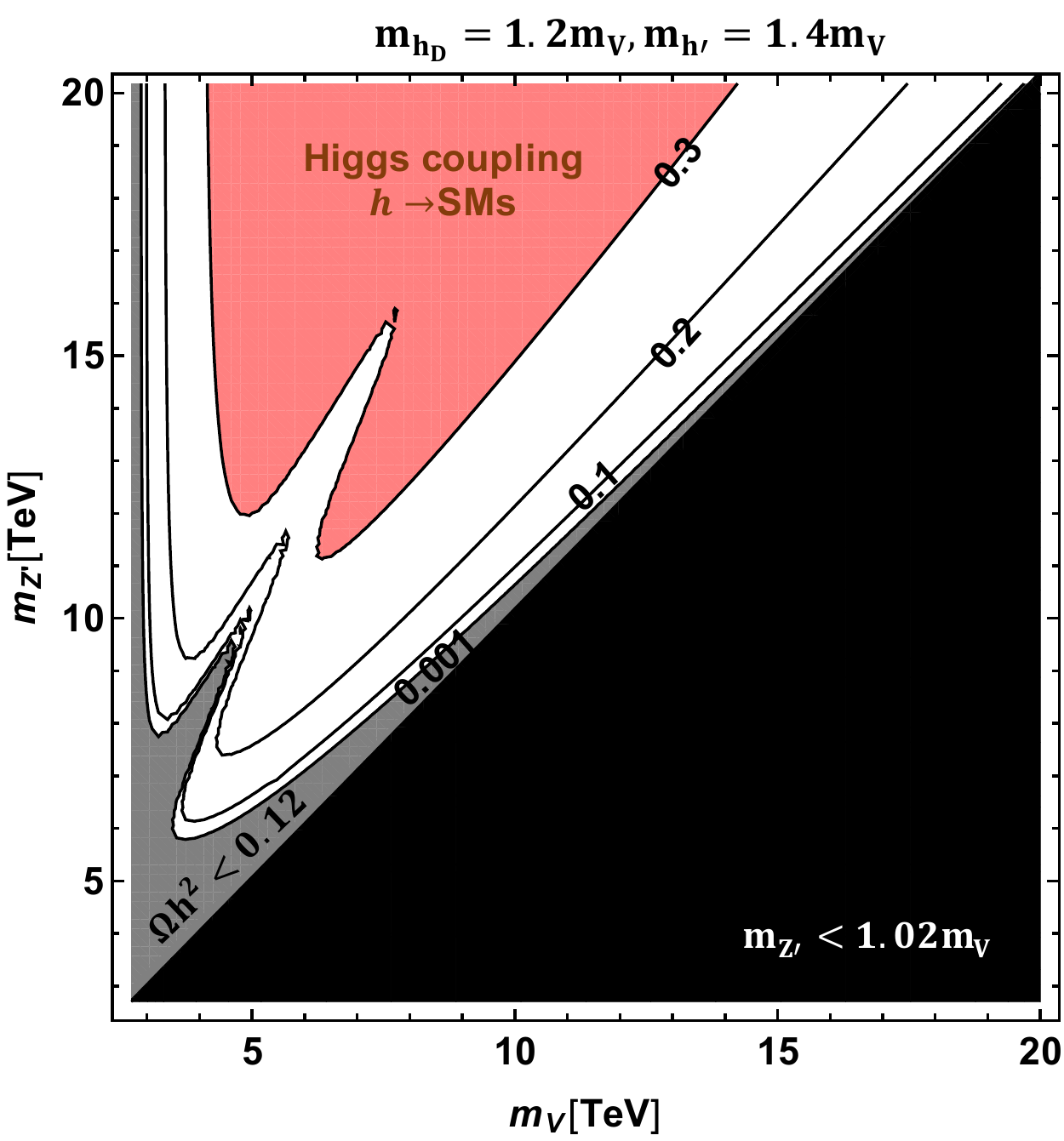}
\caption{The contours show $\phi_h$ that reproduce the measured value of the DM relic abundance. Here, 
$m_{h_D} = 1.2~m_V$ and $m_{h'} = 1.4~m_V$.
In the gray shaded region, this model cannot explain the whole abundance.
The pink region ($\phi_h > 0.3$) is constrained by the measurement of the Higgs signal strength~\cite{1909.02845}.
In the black shaded region, $g_0$ is beyond the perturbative unitarity bound.
}
\label{fig:RmVmZp}
\end{figure}

Figure~\ref{fig:RmVmZp} shows the value of $\phi_h$ that is required to obtain the measured value of the DM energy density. 
Comparing to Fig.~\ref{fig:abundance}, 
the viable region that explains the right amount of DM relic abundance is extended.
The larger $\phi_h$ requires the heavier $m_{Z'}$.
This is because $h$ and $h'$ contribute to the annihilation of pairs of DM particles for larger $\phi_h$, and the contributions of $W'$ and $Z'$ have to be smaller.
On the other hand, the region with the larger $\phi_h$ is excluded 
by the constraint on the SM Higgs couplings as we discussed in Sec.~\ref{sec:HiggsSignalStrength}. 
As a result, we can constrain the value of $m_{Z'}$ for a given $m_V$. 

\begin{figure}[tbp]
\centering
\includegraphics[width=0.46\hsize]{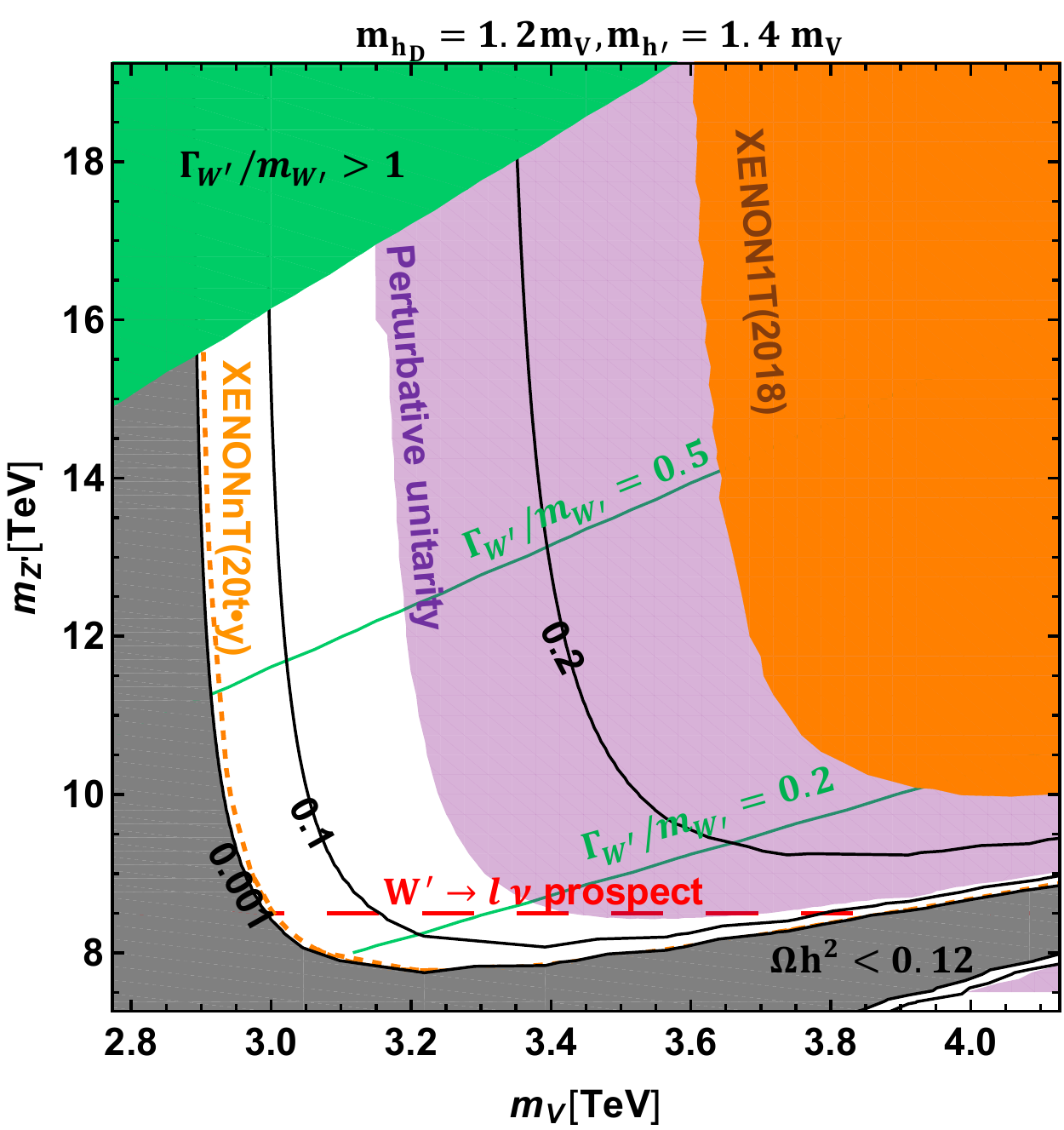}
\includegraphics[width=0.48\hsize]{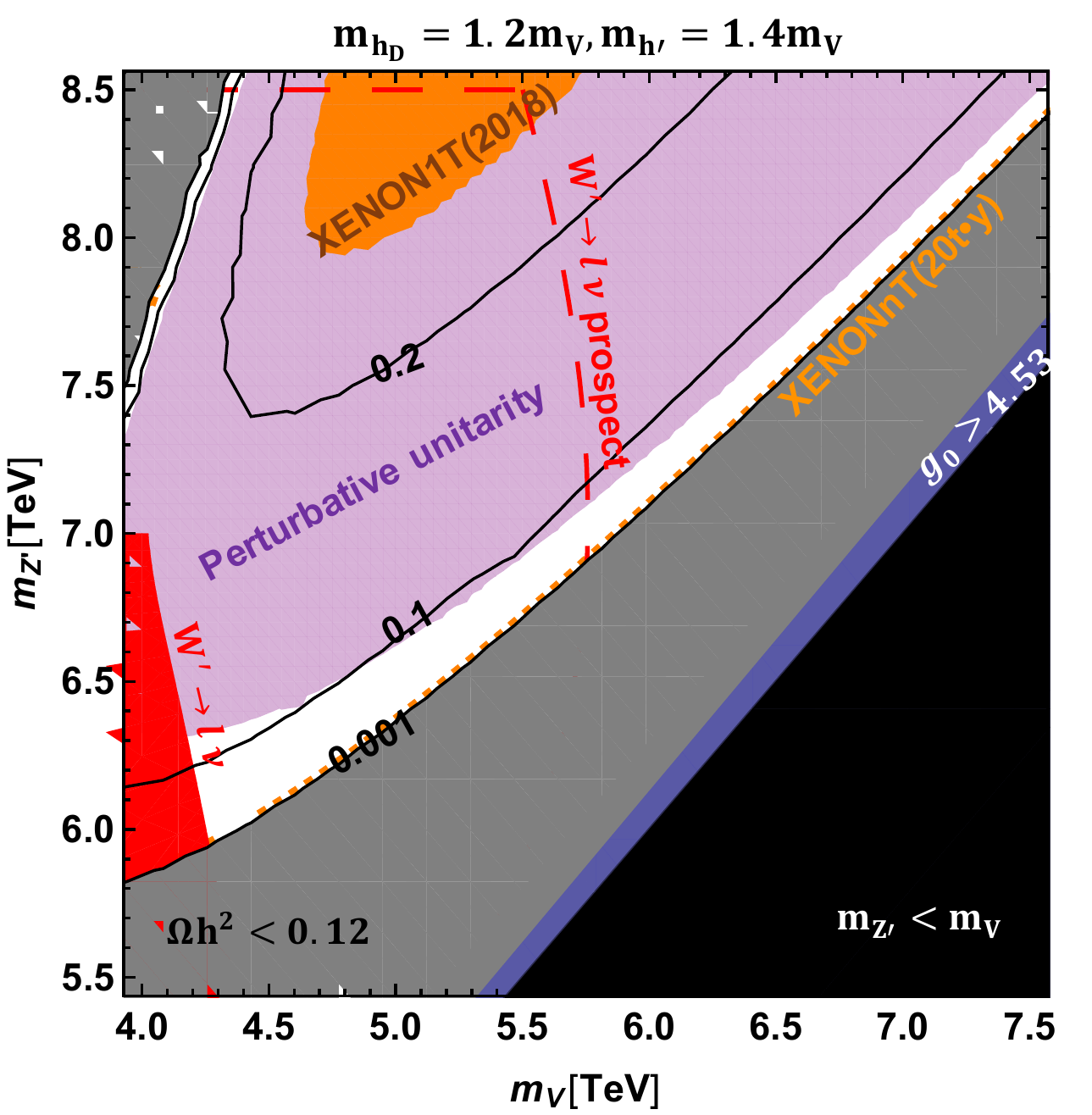}
\caption{Combined results in the $m_V$-$m_{Z'}$ plane. The gray shaded region cannot explain the whole abundance and the black shaded region is theoretically forbidden.  The orange region is already constrained by the XENON2018 and the dotted line shows the prospect by XENONnT (20~ton$\cdot$year). In the purple region, some of the Higgs quartic couplings are non-perturbative. The red shaded region is constrained by ATLAS and the red dashed line shows the prospect. 
In the left panel, the green line shows the value of $\Gamma_{W'} / m_{W'}$,
and $\Gamma_{W'} > m_{W'}$ in the green shaded region. 
In the right panel, $g_0$ is beyond the perturbative unitarity bound in the blue shaded region.
}
\label{full}
\end{figure}

We discuss the lighter and heavier $Z'$ regions in detail. 
The left panel in Fig.~\ref{full} is for the heavier $Z'$ region.
It shows that the constraint from the XENON1T experiment is stronger than 
the one from the Higgs coupling measurements.
We find that the XENONnT experiment~\cite{Aprile:2015uzo} can cover most of the parameter space for $\phi_h \gtrsim {\cal O}(0.01)$.
The constraint from the perturbative unitarity gives a stronger constraint than one from the XENON1T experiment.
However, this constraint highly depends on the choice of $m_{h'}$.
The right panel in Fig.~\ref{full} is for the lighter $Z'$ region.
The XENONnT covers the large region of the parameter space.
The HL-LHC is also useful to test the model for $m_{W'} < 5.7$~TeV.
The $W'$ search at the collider experiment is independent of $\phi_h$,
therefore the XENONnT experiment and the HL-LHC is complementary to each other.

\begin{figure}[tbp]
\centering
\begin{minipage}{1\hsize}
   \begin{minipage}[t]{0.49\hsize}
       \begin{center}
       \includegraphics[width=7cm]{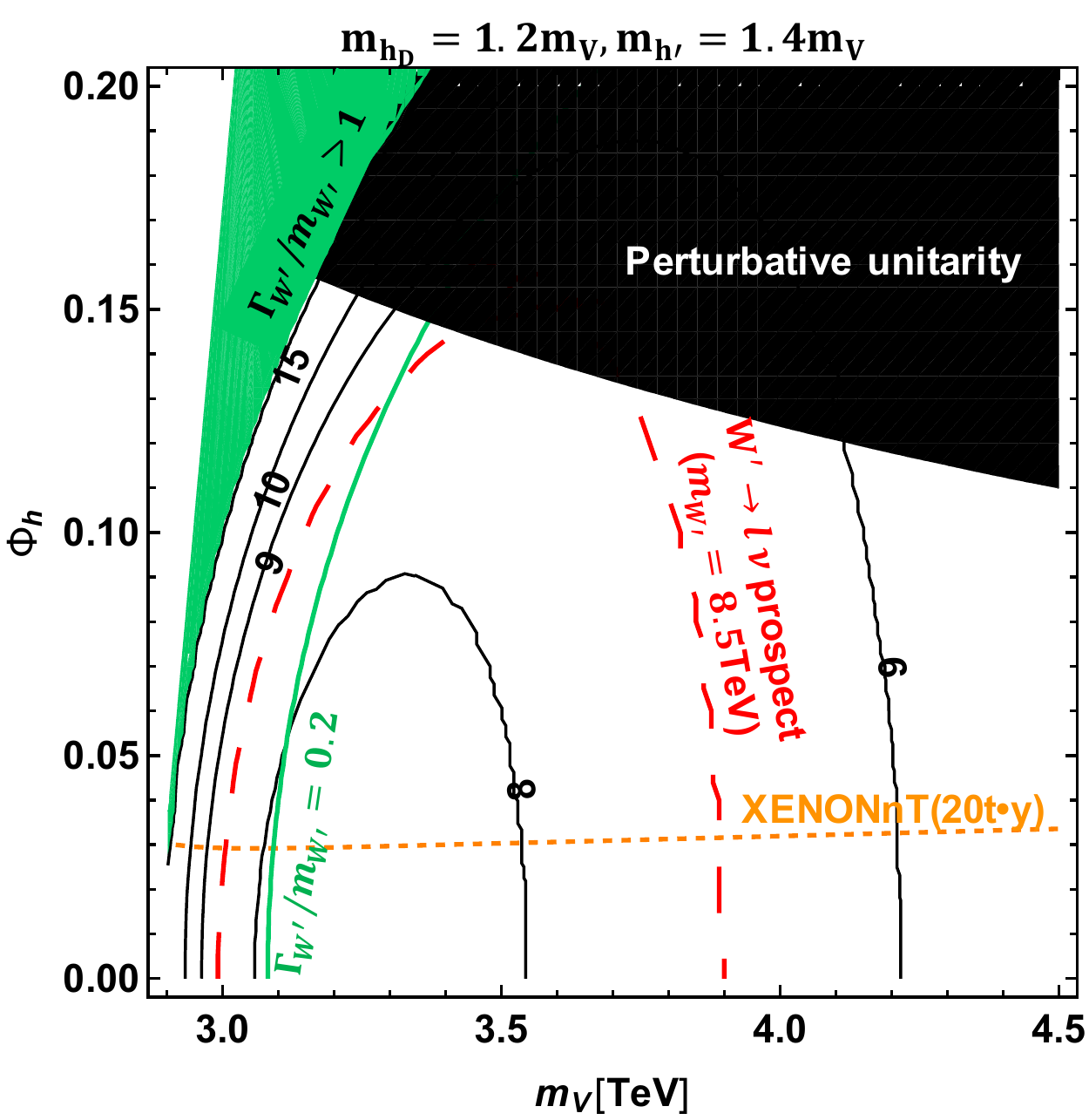}
       \end{center}
   \end{minipage}
   \begin{minipage}[t]{0.49\hsize}
       \begin{center}
       \includegraphics[width=7cm]{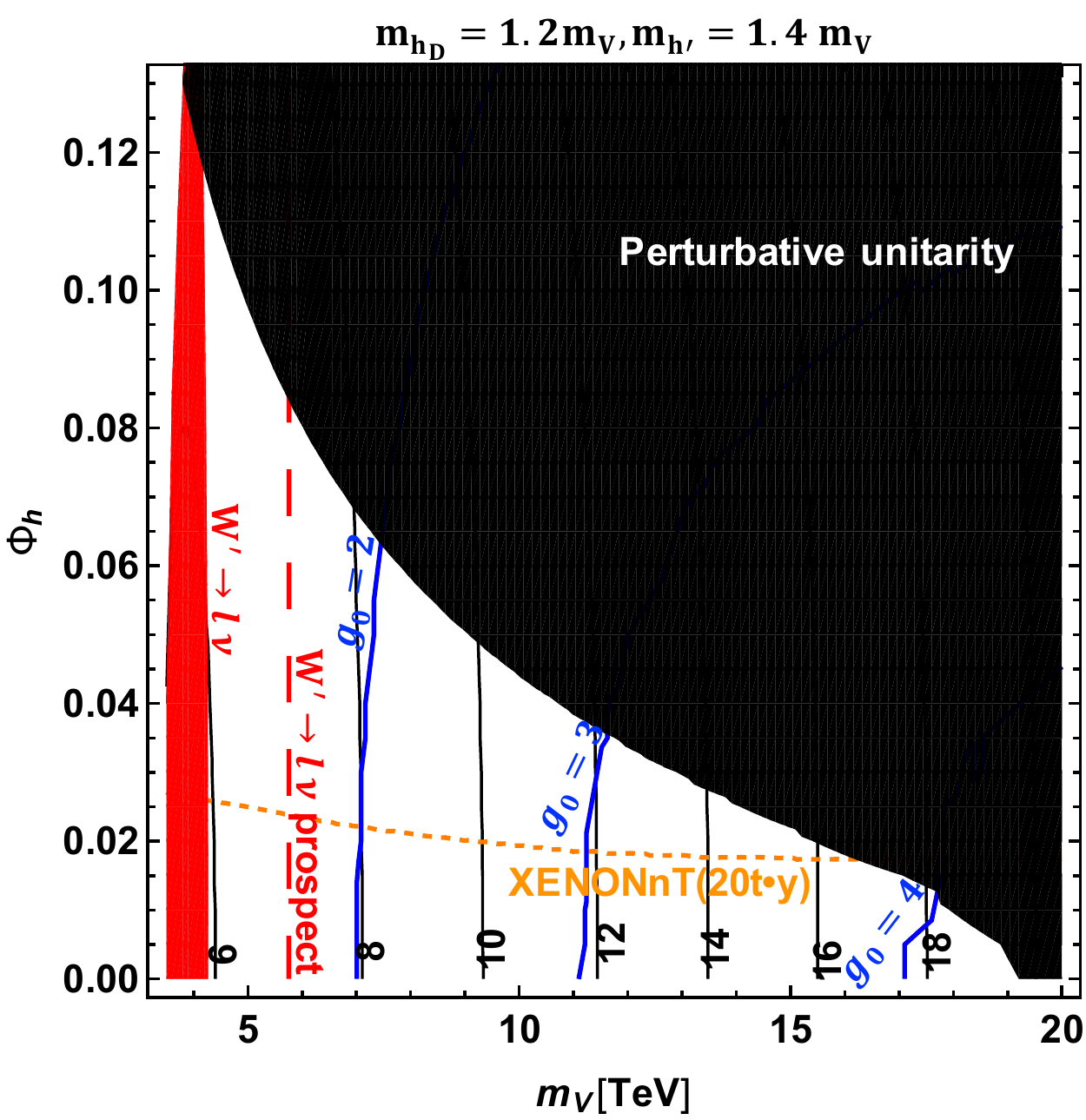}
       \end{center}
    \end{minipage}
\end{minipage}
\caption{Combined results in the $m_V$-$\phi_h$ plane.
The left (right) panel shows the case for $m_{Z'} > 2 m_V$ ($m_{Z'} < 2 m_V$).
We determine $m_{Z'}$ to obtain the right amount of the DM relic abundance, and the values are shown by the black-solid contours in TeV unit.
The orange dotted line shows the prospect of XENONnT~\cite{Aprile:2015uzo}.
The red dashed line shows the prospect of $W'$ search~\cite{ATL-PHYS-PUB-2018-044}.
In the left panel, the green line shows the value of $\Gamma_{W'} / m_{W'}$,
and $\Gamma_{W'} > m_{W'}$ in the green shaded region.
The red shaded region in the right panel is already constrained  
by the $W'$ search~\cite{1906.05609}.
The black shaded regions in both panels are excluded 
by the perturbative unitarity of the scalar quartic couplings.
The blue-solid contours in the right panel show the value of $g_0$.
}
\label{fig:RmVphiH}
\end{figure}

The smaller $\phi_h$ region is degenerate in Figs.~\ref{fig:RmVmZp} and \ref{full}. We magnify those regions in Fig.~\ref{fig:RmVphiH}.
The values of $m_{Z'}$ that are required to obtain the right amount of DM energy density are shown in the  $m_V$-$\phi_h$ plane.
The left panel shows the lighter $m_V$ region.
We find that the combination of the DM direct detection at the XENONnT experiment and the $W'$ search at the HL-LHC is a powerful tool to seek this region. 
The former will give an upper bound on $\phi_h$ that is almost independent of $m_V$. 
The latter, on the other hand, is sensitive for $3$~TeV $ \lesssim m_V \lesssim 3.9$~TeV.
For the lighter $m_V$, $m_{V} \lesssim 3$~TeV, the $W'$ decay width can be as large as $m_V$, but in most of the region it satisfies $0.1 < \Gamma_{W'}/m_{W'} < 0.2$.
The right panel in Fig.~\ref{fig:RmVphiH} is for $V$ that is heavier than $4$~TeV.
The direct detection experiment is important in this region as well to determine the value of $\phi_h$.
For $m_V \lesssim 6$~TeV, we can test this model from the $W'$ search.
We find that the perturbative unitarity of scalar quartic couplings gives the upper limit on $m_V$, $m_V \lesssim 19$~TeV. 

\section{Conclusions}
\label{sec:summary}

We constructed a model of spin-1 dark matter that has the electroweak gauge interaction.
The electroweak gauge symmetry is extended into SU(2)$_0 \times$SU(2)$_1 \times$SU(2)$_2 \times$U(1)$_Y$, and the discrete symmetry under the exchanging of SU(2)$_0$ and SU(2)$_2$ is imposed. 
It is not necessary to extend the fermion sector to realize the realistic fermion mass spectra through the Yukawa interactions. Since the dark matter candidate in this model couples to the electroweak gauge bosons, we do not need to rely on the Higgs portal couplings. These two features are distinctive of our model from other spin-1 dark matter models.
Our model predicts spin-0 and spin-1 dark matter candidates, and the heavier one decays into the lighter one. In this paper, we focus on the spin-1 dark matter candidate.

The model predicts a heavy vector triplet ($W'^\pm$ and $Z'$) in the visible sector. We found that the $W'$ searches at the LHC give a strong constraint.
That has already excluded some regions of the parameter space that can explain the measured value of the dark matter energy density by the freeze-out mechanism.

There are three scenarios that the model predicts the right amount of the dark matter relic abundance.
The first scenario is that the heavy vector triplet is slightly heavier than the dark matter but has almost degenerate mass. In this case, pairs of dark matter particles can annihilate into a heavy triplet and a SM particle. This process is efficient, and the measured value of dark matter energy density is explained for $m_V \gtrsim 4$~TeV. 
The upper bound on the mass of the dark matter is imposed by the perturbative unitarity bound of the gauge couplings, $m_V \lesssim 19$~TeV.
The HL-LHC can test this scenario up to 6~TeV. 
The second scenario is for $m_{W'} \simeq 2 m_V$ that utilizes the $W'$ resonance in the (co)annihilation processes of pairs of dark matter particles. In this case, the gauge couplings are well in the perturbative regime.
The third scenario is for $m_{W'} \gg m_V$. 
In this scenario, the mass of the dark matter is almost uniquely determined with the assumption that the relic abundance explains the full of the dark matter energy density, $m_V \simeq 3$~TeV. This last scenario is similar to other SU(2)$_L$-triplet dark matter models. The mass of the $W'$ is bounded by the condition that $\Gamma_{W'} < m_{W'}$, and we find $m_{W'} \lesssim$ 15~TeV in the small scalar mixing limit.

Although we do not need to rely on the Higgs portal interactions in this model, it predicts the signal for the direct detection experiments, and thus we also discussed the effects of the scalar mixing.
We found that the perturbative unitarity bounds for the scalar quartic couplings give a stronger constraint on the mixing. We also found that the model is testable at the XENONnT experiment if $|\phi_h| \gtrsim 0.03$.

Since our dark matter interacts with the electroweak gauge bosons and is much heavier than them, the Sommerfeld enhancement is expected to give significant effects~\cite{hep-ph/0212022, hep-ph/0307216, hep-ph/0412403, 0810.0713, 1603.01383}. It may alter our results for the relic abundance.
The effect is also important to test this model by the indirect detection experiments. 
We leave this to further study.

\section*{Acknowledgments}
This work was supported by JSPS KAKENHI Grant Number 16K17715 and 19H04615 [T.A.],
and by Grant-in-Aid for Scientific research from the Ministry of
Education, Science, Sports, and Culture (MEXT), Japan, No. 16H06492
[J.H.]. The work is also supported by World Premier
International Research Center Initiative (WPI Initiative), MEXT,
Japan [J.H.] and by JSPS Core-to-Core Program (grant number:JPJSCCA20200002).

\appendix
\section{Some details in the gauge sectors}
\label{app:gauge-sector}
The mass eigenstates are given by
\begin{align}
 \begin{pmatrix}
  V^{\pm}_\mu \\  W^{\pm}_\mu \\ W'^{\pm}_{\mu}
 \end{pmatrix}
=
\begin{pmatrix}
 \omega^0_V & \omega^1_V & \omega^2_V\\
 \omega^0_{W} & \omega^1_W & \omega^2_W\\
 \omega^0_{W'} & \omega^1_{W'} & \omega^2_{W'}\\
\end{pmatrix}
 \begin{pmatrix}
  W_{0\mu}^{\pm} \\ W_{1\mu}^{\pm} \\ W_{2\mu}^{\mu}
 \end{pmatrix}
=
\begin{pmatrix}
 \frac{1}{\sqrt{2}} & 0 & -\frac{1}{\sqrt{2}}\\
 \frac{\sin\phi_\pm}{\sqrt{2}} & \cos\phi_\pm & \frac{\sin\phi_\pm}{\sqrt{2}}\\
 \frac{\cos\phi_\pm}{\sqrt{2}} & -\sin\phi_\pm & \frac{\cos\phi_\pm}{\sqrt{2}}\\
\end{pmatrix}
 \begin{pmatrix}
  W_{0\mu}^{\pm} \\ W_{1\mu}^{\pm} \\ W_{2\mu}^{\mu}
 \end{pmatrix}
,\\
 \begin{pmatrix}
  V^{0}_\mu \\  A_\mu \\ Z_\mu \\ Z'^{\pm}_{\mu}
 \end{pmatrix}
=
\begin{pmatrix}
 \omega^0_V & \omega^1_V & \omega^2_V & \omega^B_V \\
 \omega^0_{\gamma} & \omega^1_\gamma & \omega^2_\gamma & \omega^B_\gamma\\
 \omega^0_{Z} & \omega^1_Z & \omega^2_Z & \omega^B_Z\\
 \omega^0_{Z'} & \omega^1_{Z'} & \omega^2_{Z'} & \omega^B_{Z'}\\
\end{pmatrix}
 \begin{pmatrix}
  W_{0\mu}^{3} \\ W_{1\mu}^{3} \\ W_{2\mu}^{3} \\ B_\mu
 \end{pmatrix}
=
\begin{pmatrix}
 \frac{1}{\sqrt{2}} & 0 & - \frac{1}{\sqrt{2}} & 0 \\
 \frac{e}{g_0} &  \frac{e}{g_1} &  \frac{e}{g_0} &  \frac{e}{g'} \\
 \omega^0_{Z} & \omega^1_Z & \omega^0_Z & \omega^B_Z\\
 \omega^0_{Z'} & \omega^1_{Z'} & \omega^0_{Z'} & \omega^B_{Z'}\\
\end{pmatrix}
 \begin{pmatrix}
  W_{0\mu}^{3} \\ W_{1\mu}^{3} \\ W_{2\mu}^{3} \\ B_\mu
 \end{pmatrix}
,
\end{align}
where
\begin{align}
 e =& \left( \frac{2}{g_0^2} + \frac{1}{g_1^2} + \frac{1}{g'^2} \right)^{-1/2},
\\
\omega^0_Z = \omega^2_Z
=& 
\frac{e g_1}{\sqrt{g_0^2 + 2 g_1^2} g'} \cos\phi_0
+
\frac{g_0}{\sqrt{2(g_0^2 + 2 g_1^2)}} \sin\phi_0
,\\
\omega^1_Z
=& 
\frac{e g_0}{\sqrt{g_0^2 + 2 g_1^2} g'} \cos\phi_0
-
\frac{\sqrt{2} g_1}{\sqrt{g_0^2 + 2 g_1^2}} \sin\phi_0
,\\
\omega^B_Z
=& 
-\frac{e \sqrt{g_0^2 + 2 g_1^2}}{g_0 g_1} \cos\phi_0
,\\
\omega^0_{Z'} = \omega^2_{Z'}
=& 
\frac{g_0}{\sqrt{2(g_0^2 + 2 g_1^2)}} \cos\phi_0
-
\frac{e g_1}{\sqrt{g_0^2 + 2 g_1^2} g'} \sin\phi_0
,\\
\omega^1_{Z'}
=& 
-\frac{\sqrt{2} g_1}{\sqrt{g_0^2 + 2 g_1^2}} \cos\phi_0
-\frac{e g_0}{\sqrt{g_0^2 + 2 g_1^2} g'} \sin\phi_0
,\\
\omega^B_{Z'}
=& 
\frac{e \sqrt{g_0^2 + 2 g_1^2}}{g_0 g_1} \sin\phi_0
.
\end{align}
Here we introduce $\phi_\pm$ and $\phi_0$ that satisfy
\begin{align}
\frac{1}{4}
 \begin{pmatrix}
  g_1^2 (v^2 + 2 v_\Phi^2)   &  -\sqrt{2} g_0 g_1 v_\Phi^2 \\  
  -\sqrt{2} g_0 g_1 v_\Phi^2 &   g_0^2 v_\Phi^2 
 \end{pmatrix}
\begin{pmatrix}
  \cos\phi_\pm &  -\sin\phi_\pm \\ 
 \sin\phi_\pm &  \cos\phi_\pm 
\end{pmatrix}
=&
\begin{pmatrix}
 \cos\phi_\pm &  -\sin\phi_\pm \\ 
 \sin\phi_\pm &  \cos\phi_\pm 
\end{pmatrix}
 \begin{pmatrix}
  m_W^2 & 0 \\ 0 & m_{W'}^2
 \end{pmatrix}
,\\
\frac{1}{4}\begin{pmatrix}
  \frac{g_0^2 g_1^2 g'^2}{e^2 (g_0^2 + 2 g_1^2)} v^2   &  -\frac{\sqrt{2} g_0 g_1^3 g'}{e (g_0^2 + 2 g_1^2)} v^2 \\  
  -\frac{\sqrt{2} g_0 g_1^3 g'}{e (g_0^2 + 2 g_1^2)} v^2 &  \frac{(g_0^2 + 2 g_1^2)^2 v_\Phi^2 + 2 g_1^4 v^2}{(g_0^2 + 2 g_1^2)}
 \end{pmatrix}
\begin{pmatrix}
  \cos\phi_0 &  -\sin\phi_0 \\ 
 \sin\phi_0 &  \cos\phi_0 
\end{pmatrix}
=&
\begin{pmatrix}
 \cos\phi_0 &  -\sin\phi_0 \\ 
 \sin\phi_0 &  \cos\phi_0 
\end{pmatrix}
 \begin{pmatrix}
  m_Z^2 & 0 \\ 0 & m_{Z'}^2
 \end{pmatrix}
.
\end{align}
We find
\begin{align}
 \cos^2\phi_\pm = \frac{m_{V^\pm}^2 - m_W^2}{m_{W'}^2 - m_W^2}.
\end{align}
One can always choose $\cos\phi_\pm > 0$ as a convention, and thus
\begin{align}
 \cos\phi_\pm = \sqrt{\frac{m_{V^\pm}^2 - m_W^2}{m_{W'}^2 - m_W^2}},
\quad
 \sin\phi_\pm = \sqrt{\frac{m_{W'}^2 - m_{V^\pm}^2}{m_{W'}^2 - m_W^2}}.
\end{align}
We also find
\begin{align}
 m_W^2
=&
\frac{1}{8}
\left\{
g_1^2 v^2 + (g_0^2 + 2 g_1^2) v_\Phi^2
-
\sqrt{
-4 g_0^2 g_1^2 v^2 v_\Phi^2
+
\left[ g_1^2 v^2 + (g_0^2 + 2 g_1^2 ) v_\Phi^2 \right]^2
}
\right\}
,\\
 m_{W'}^2
=&
\frac{1}{8}
\left\{
g_1^2 v^2 + (g_0^2 + 2 g_1^2) v_\Phi^2
+
\sqrt{
-4 g_0^2 g_1^2 v^2 v_\Phi^2
+
\left[ g_1^2 v^2 + (g_0^2 + 2 g_1^2 ) v_\Phi^2 \right]^2
}
\right\}
.
\end{align}
For $v_\Phi \gg v$, the mixing angles are given by
\begin{align}
\cos\phi_\pm =& \frac{g_0}{\sqrt{g_0^2 + 2g_1^2}} + {\cal O}(v_\Phi^{-2})
\simeq \frac{m_V}{m_{Z'}}
,\\
\sin\phi_\pm =& \frac{\sqrt{2} g_1}{\sqrt{g_0^2 + 2g_1^2}} + {\cal O}(v_\Phi^{-2})
\simeq \sqrt{1 - \frac{m_V^2}{m_{Z'}^2}}
,\\
\cos\phi_0 =& 1 + {\cal O}(v_\Phi^{-4}),\\
\sin\phi_0 =& \frac{\sqrt{2} g_0 g_1^3 g'}{e(g_0^2 + 2 g_1^2)^2} \frac{v^2}{v_\Phi^2} + {\cal O}(v_\Phi^{-4}).
\end{align}

\section{Would-be NG bosons}
\label{app:NG-sector}
The mass matrices for the gauge bosons are given by
\begin{align}
 {\cal M}_C^2 = Q_W^t Q_W,
\quad
 {\cal M}_N^2 = Q_Z^t Q_Z,
\end{align}
where
\begin{align}
Q_W =&
\frac{1}{2}
\begin{pmatrix}
 g_0 v_\Phi & -g_1 v_\Phi & 0 \\
 0 & g_1 v & 0 & \\
 0 & -g_1 v_\Phi & g_0 v_\Phi \\
\end{pmatrix}
,\\
Q_Z=&
\frac{1}{2}
\begin{pmatrix}
 g_0 v_\Phi & -g_1 v_\Phi & 0 & 0 \\ 
 0 & g_1 v & 0 & -g' v\\
 0 & -g_1 v_\Phi & g_0 v_\Phi & 0 
\end{pmatrix}
.
\end{align}
In the $R_\xi$ gauge, the mass terms are given by
\begin{align}
-\begin{pmatrix}
 \pi_1^+ &  \pi_3^+ & \pi_2^+        
\end{pmatrix}
\xi Q_W Q_W^t
\begin{pmatrix}
 \pi_1^- \\  \pi_3^+  \\  \pi_2^- 
\end{pmatrix}
- \frac{1}{2}
\begin{pmatrix}
 \pi_1^0 &  \pi_3^0  &  \pi_2^0 
\end{pmatrix}
\xi Q_Z Q_Z^t
\begin{pmatrix}
 \pi_1^0 \\  \pi_3^0 \\  \pi_2^0 
\end{pmatrix}
.
\end{align}
The eigenvectors of the mass matrices are
\begin{align}
 Q_W^t Q_W \vec{\omega}_X =& m_X^2 \vec{\omega}_X, \label{eq:QQwW}\\
 Q_W Q_W^t \vec{\omega}_{\pi_X}  =& m_X^2 \vec{\omega}_{\pi_X},\label{eq:QQwpi}\\ 
 Q_Z^t Q_Z \vec{\omega}_X =& m_X^2 \vec{\omega}_X, \\
 Q_Z Q_Z^t \vec{\omega}_{\pi_X}  =& m_X^2 \vec{\omega}_{\pi_X}. 
\end{align}
Multiplying $Q_W$ to Eq.~\eqref{eq:QQwW} and comparing it with Eq.~\eqref{eq:QQwpi},
one can find that $Q_W \vec{\omega}_X \propto \vec{\omega}_{\pi_X}$. 
Note that $(Q_W \vec{\omega}_X)^t (Q_W \vec{\omega}_X) = 
\vec{\omega}_X^t Q_W^t Q_W \vec{\omega}_X = m_X^2$
and $\vec{\omega}_{\pi_X}^t \vec{\omega}_{\pi_X} =1$. 
Similar relations are also found in the neutral sector. 
Finally, we find
\begin{align}
 Q_W \vec{\omega}_X =& m_X \vec{\omega}_{\pi_X}, \label{eq:Charged-Gauge-and-NG-mixing}\\
 Q_Z \vec{\omega}_X =& m_X \vec{\omega}_{\pi_X}. \label{eq:Neutral-Gauge-and-NG-mixing}
\end{align}
These relations are useful to obtain the Fermi constant
and some relation among couplings. 
For example, we use 
\begin{align}
  g_1 v \omega_X^1 =& 2 m_X \omega_{\pi_X}^3
\end{align}
to obtain the Fermi constant.

The mixing angles for the charged NG-bosons are given by
\begin{align}
 \begin{pmatrix}
  \vec{\omega}_{\pi_{V^\pm}} &   \vec{\omega}_{\pi_{W}}  &   \vec{\omega}_{\pi_{W'}} 
 \end{pmatrix}
=
 \begin{pmatrix}
  \omega_{\pi_{V^\pm}}^1 &   \omega_{\pi_{W}}^1  &  \omega_{\pi_{W'}}^1 \\ 
  \omega_{\pi_{V^\pm}}^3 &   \omega_{\pi_{W}}^3  &  \omega_{\pi_{W'}}^3 \\ 
  \omega_{\pi_{V^\pm}}^2 &   \omega_{\pi_{W}}^2  &  \omega_{\pi_{W'}}^2 \\ 
 \end{pmatrix}
=
 \begin{pmatrix}
  \frac{1}{\sqrt{2}} & \frac{\sin\phi_\pi}{\sqrt{2}} & \frac{\cos\phi_\pi}{\sqrt{2}} \\
  0                  & \cos\phi_\pi & -\sin\phi_\pi \\
 -\frac{1}{\sqrt{2}} & \frac{\sin\phi_\pi}{\sqrt{2}} & \frac{\cos\phi_\pi}{\sqrt{2}} \\
 \end{pmatrix}
,
\label{eq:charged-NG-mixing}
\end{align}
where $\sin\phi_\pi$ and $\cos\phi_\pi$ satisfy
\begin{align}
\frac{1}{4}
 \begin{pmatrix}
  g_1^2 v^2   &  -\sqrt{2} g_1^2 v v_\Phi \\  
  -\sqrt{2} g_1^2 v v_\Phi &  (g_0^2 + 2 g_1^2) v_\Phi^2 
 \end{pmatrix}
\begin{pmatrix}
 \cos\phi_\pi &  -\sin\phi_\pi \\ 
 \sin\phi_\pi &  \cos\phi_\pi
\end{pmatrix}
=
\begin{pmatrix}
 \cos\phi_\pi &  -\sin\phi_\pi \\ 
 \sin\phi_\pi &  \cos\phi_\pi 
\end{pmatrix}
 \begin{pmatrix}
  m_W^2 & 0 \\ 0 & m_{W'}^2
 \end{pmatrix}
.
\end{align}
Comparing Eqs.~\eqref{eq:charged-NG-mixing} and \eqref{eq:Charged-Gauge-and-NG-mixing},
we find
\begin{align}
 \cos\phi_\pi =& \frac{g_1 v}{2 m_W} \cos\phi_\pm,\\
 \sin\phi_\pi =& \frac{g_1 v}{2 m_{W'}} \sin\phi_\pm.
\end{align}

The mixing angles for the neutral NG-bosons are given by
\begin{align}
 \begin{pmatrix}
  \vec{\omega}_{\pi_{V^0}} &   \vec{\omega}_{\pi_{Z}}  &   \vec{\omega}_{\pi_{Z'}} 
 \end{pmatrix}
=
 \begin{pmatrix}
  \omega_{\pi_{V^0}}^1 &   \omega_{\pi_{Z}}^1  &  \omega_{\pi_{Z'}}^1 \\ 
  \omega_{\pi_{V^0}}^3 &   \omega_{\pi_{Z}}^3  &  \omega_{\pi_{Z'}}^3 \\ 
  \omega_{\pi_{V^0}}^2 &   \omega_{\pi_{Z}}^2  &  \omega_{\pi_{Z'}}^2 \\ 
 \end{pmatrix}
=
 \begin{pmatrix}
  \frac{1}{\sqrt{2}} & \frac{\sin\phi_{\pi_0}}{\sqrt{2}} & \frac{\cos\phi_{\pi_0}}{\sqrt{2}} \\
  0                  & \cos\phi_{\pi_0} & -\sin\phi_{\pi_0} \\
  -\frac{1}{\sqrt{2}} & \frac{\sin\phi_{\pi_0}}{\sqrt{2}} & \frac{\cos\phi_{\pi_0}}{\sqrt{2}} \\
 \end{pmatrix}
,
\label{eq:neutral-NG-mixing}
\end{align}
where $\sin\phi_{\pi_0}$ and $\cos\phi_{\pi_0}$ satisfy
\begin{align}
\frac{1}{4}
 \begin{pmatrix}
  (g_1^2 + g'^2) v^2   &  -\sqrt{2} g_1^2 v v_\Phi \\  
  -\sqrt{2} g_1^2 v v_\Phi &  (g_0^2 + 2 g_1^2) v_\Phi^2 
 \end{pmatrix}
\begin{pmatrix}
 \cos\phi_{\pi_0} &  -\sin\phi_{\pi_0} \\ 
 \sin\phi_{\pi_0} &  \cos\phi_{\pi_0}
\end{pmatrix}
=
\begin{pmatrix}
 \cos\phi_{\pi_0} &  -\sin\phi_{\pi_0} \\ 
 \sin\phi_{\pi_0} &  \cos\phi_{\pi_0} 
\end{pmatrix}
 \begin{pmatrix}
  m_Z^2 & 0 \\ 0 & m_{Z'}^2
 \end{pmatrix}
.
\end{align}
Comparing Eqs.~\eqref{eq:neutral-NG-mixing} and \eqref{eq:Neutral-Gauge-and-NG-mixing},
we find
\begin{align}
 \cos\phi_{\pi_0} =& \frac{v}{2 m_{Z}} (g_1 \omega^1_Z - g' \omega^B_Z),\\
 \sin\phi_{\pi_0} =& -\frac{v}{2 m_{Z'}} (g_1 \omega^1_{Z'} - g' \omega^B_{Z'}).
\end{align}

\section{Fermi constant}
\label{app:fermi-constant}
The Fermi constant is defined by the muon decay, $\mu \to \nu_\mu e \bar{\nu}_{e}$.
There is a $W'$ exchanging diagram as well as the $W$-exchanging diagram. 
We have to add both contributions.
We can simplify the calculation by using the relation between the mixing angles in the gauge sector and NG-boson sector.
The Fermi constant is given by
\begin{align}
\sqrt{2} G_F \equiv 
&
\sum_{X=W,W'} \frac{g_{X \bar{\nu_\mu}} g_{X \bar{e} \nu_e}}{4 m_X^2}
\nonumber\\
=&
\sum_{X=W,V,W'} \frac{(g_1 \omega^1_X)^2}{4 m_X^2}
\nonumber\\
=&
\sum_{X=W,V,W'} \frac{(\omega_{\pi_X}^3)^2}{v^2 }
\nonumber\\
=&
\frac{1}{v^2}
.
\end{align}
In the last line, we used that $\sum_X \omega_{\pi_X}^j \omega_{\pi_X}^k = \delta^{jk}$.
Therefore, we find
\begin{align}
 \sqrt{2} G_F = \frac{1}{v^2}.
\end{align}


\begin{thebibliography}{99}
\bibitem{1807.06209} 
  N.~Aghanim {\it et al.} [Planck Collaboration],
  arXiv:1807.06209 [astro-ph.CO].


\bibitem{Lee:1977ua} 
  B.~W.~Lee and S.~Weinberg,
  Phys.\ Rev.\ Lett.\  {\bf 39}, 165 (1977).
  doi:10.1103/PhysRevLett.39.165


\bibitem{1805.12562} 
  E.~Aprile {\it et al.} [XENON Collaboration],
  Phys.\ Rev.\ Lett.\  {\bf 121}, no. 11, 111302 (2018)
  doi:10.1103/PhysRevLett.121.111302
  [arXiv:1805.12562 [astro-ph.CO]].


\bibitem{1404.3716} 
  S.~Ipek, D.~McKeen and A.~E.~Nelson,
  Phys.\ Rev.\ D {\bf 90}, no. 5, 055021 (2014)
  doi:10.1103/PhysRevD.90.055021
  [arXiv:1404.3716 [hep-ph]].


\bibitem{1408.4929} 
  K.~Ghorbani,
  JCAP {\bf 1501}, 015 (2015)
  doi:10.1088/1475-7516/2015/01/015
  [arXiv:1408.4929 [hep-ph]].


\bibitem{1701.04131} 
  S.~Baek, P.~Ko and J.~Li,
  Phys.\ Rev.\ D {\bf 95}, no. 7, 075011 (2017)
  doi:10.1103/PhysRevD.95.075011
  [arXiv:1701.04131 [hep-ph]].


\bibitem{1708.02253} 
  C.~Gross, O.~Lebedev and T.~Toma,
  Phys.\ Rev.\ Lett.\  {\bf 119}, no. 19, 191801 (2017)
  doi:10.1103/PhysRevLett.119.191801
  [arXiv:1708.02253 [hep-ph]].


\bibitem{Abe:2020iph} 
  Y.~Abe, T.~Toma and K.~Tsumura,
  JHEP {\bf 2005}, 057 (2020)
  doi:10.1007/JHEP05(2020)057
  [arXiv:2001.03954 [hep-ph]].


\bibitem{Okada:2020zxo} 
  N.~Okada, D.~Raut and Q.~Shafi,
  arXiv:2001.05910 [hep-ph].


\bibitem{Ahmed:2020hiw} 
  A.~Ahmed, S.~Najjari and C.~B.~Verhaaren,
  JHEP {\bf 2006}, 007 (2020)
  doi:10.1007/JHEP06(2020)007
  [arXiv:2003.08947 [hep-ph]].


\bibitem{Maru:2018ocf} 
  N.~Maru, N.~Okada and S.~Okada,
  Phys.\ Rev.\ D {\bf 98}, no. 7, 075021 (2018)
  doi:10.1103/PhysRevD.98.075021
  [arXiv:1803.01274 [hep-ph]].


\bibitem{Belyaev:2018xpf} 
  A.~Belyaev, G.~Cacciapaglia, J.~Mckay, D.~Marin and A.~R.~Zerwekh,
  Phys.\ Rev.\ D {\bf 99}, no. 11, 115003 (2019)
  doi:10.1103/PhysRevD.99.115003
  [arXiv:1808.10464 [hep-ph]].


\bibitem{hep-ph/0206071} 
  G.~Servant and T.~M.~P.~Tait,
  Nucl.\ Phys.\ B {\bf 650}, 391 (2003)
  doi:10.1016/S0550-3213(02)01012-X
  [hep-ph/0206071].


\bibitem{1005.5651} 
  S.~Kanemura, S.~Matsumoto, T.~Nabeshima and N.~Okada,
  Phys.\ Rev.\ D {\bf 82}, 055026 (2010)
  doi:10.1103/PhysRevD.82.055026
  [arXiv:1005.5651 [hep-ph]].


\bibitem{1111.4482} 
  O.~Lebedev, H.~M.~Lee and Y.~Mambrini,
  Phys.\ Lett.\ B {\bf 707}, 570 (2012)
  doi:10.1016/j.physletb.2012.01.029
  [arXiv:1111.4482 [hep-ph]].


\bibitem{1202.5902} 
  T.~Abe, M.~Kakizaki, S.~Matsumoto and O.~Seto,
  Phys.\ Lett.\ B {\bf 713}, 211 (2012)
  doi:10.1016/j.physletb.2012.05.051
  [arXiv:1202.5902 [hep-ph]].


\bibitem{1207.4272} 
  Y.~Farzan and A.~R.~Akbarieh,
  JCAP {\bf 1210}, 026 (2012)
  doi:10.1088/1475-7516/2012/10/026
  [arXiv:1207.4272 [hep-ph]].


\bibitem{1212.2131} 
  S.~Baek, P.~Ko, W.~I.~Park and E.~Senaha,
  JHEP {\bf 1305}, 036 (2013)
  doi:10.1007/JHEP05(2013)036
  [arXiv:1212.2131 [hep-ph]].


\bibitem{1312.4573} 
  J.~M.~Hyde, A.~J.~Long and T.~Vachaspati,
  Phys.\ Rev.\ D {\bf 89}, 065031 (2014)
  doi:10.1103/PhysRevD.89.065031
  [arXiv:1312.4573 [hep-ph]].


\bibitem{1404.5257} 
  P.~Ko, W.~I.~Park and Y.~Tang,
  JCAP {\bf 1409}, 013 (2014)
  doi:10.1088/1475-7516/2014/09/013
  [arXiv:1404.5257 [hep-ph]].


\bibitem{1405.3530} 
  S.~Baek, P.~Ko and W.~I.~Park,
  Phys.\ Rev.\ D {\bf 90}, no. 5, 055014 (2014)
  doi:10.1103/PhysRevD.90.055014
  [arXiv:1405.3530 [hep-ph]].


\bibitem{1409.3227} 
  J.~H.~Yu,
  Phys.\ Rev.\ D {\bf 90}, no. 9, 095010 (2014)
  doi:10.1103/PhysRevD.90.095010
  [arXiv:1409.3227 [hep-ph]].


\bibitem{1410.0918} 
  C.~R.~Chen, Y.~K.~Chu and H.~C.~Tsai,
  Phys.\ Lett.\ B {\bf 741}, 205 (2015)
  doi:10.1016/j.physletb.2014.12.043
  [arXiv:1410.0918 [hep-ph]].


\bibitem{0811.0172} 
  T.~Hambye,
  JHEP {\bf 0901}, 028 (2009)
  doi:10.1088/1126-6708/2009/01/028
  [arXiv:0811.0172 [hep-ph]].


\bibitem{0910.2831} 
  H.~Zhang, C.~S.~Li, Q.~H.~Cao and Z.~Li,
  Phys.\ Rev.\ D {\bf 82}, 075003 (2010)
  doi:10.1103/PhysRevD.82.075003
  [arXiv:0910.2831 [hep-ph]].


\bibitem{1007.2631} 
  J.~L.~Diaz-Cruz and E.~Ma,
  Phys.\ Lett.\ B {\bf 695}, 264 (2011)
  doi:10.1016/j.physletb.2010.11.039
  [arXiv:1007.2631 [hep-ph]].


\bibitem{1107.2093} 
  S.~Bhattacharya, J.~L.~Diaz-Cruz, E.~Ma and D.~Wegman,
  Phys.\ Rev.\ D {\bf 85}, 055008 (2012)
  doi:10.1103/PhysRevD.85.055008
  [arXiv:1107.2093 [hep-ph]].


\bibitem{1306.2329} 
  T.~Hambye and A.~Strumia,
  Phys.\ Rev.\ D {\bf 88}, 055022 (2013)
  doi:10.1103/PhysRevD.88.055022
  [arXiv:1306.2329 [hep-ph]].


\bibitem{1309.6640} 
  H.~Davoudiasl and I.~M.~Lewis,
  Phys.\ Rev.\ D {\bf 89}, no. 5, 055026 (2014)
  doi:10.1103/PhysRevD.89.055026
  [arXiv:1309.6640 [hep-ph]].


\bibitem{Baek:2013dwa} 
  S.~Baek, P.~Ko and W.~I.~Park,
  JCAP {\bf 1410}, 067 (2014)
  doi:10.1088/1475-7516/2014/10/067
  [arXiv:1311.1035 [hep-ph]].


\bibitem{1406.2291} 
  V.~V.~Khoze and G.~Ro,
  JHEP {\bf 1410}, 061 (2014)
  doi:10.1007/JHEP10(2014)061
  [arXiv:1406.2291 [hep-ph]].


\bibitem{1409.1162} 
  S.~Fraser, E.~Ma and M.~Zakeri,
  Int.\ J.\ Mod.\ Phys.\ A {\bf 30}, no. 03, 1550018 (2015)
  doi:10.1142/S0217751X15500189
  [arXiv:1409.1162 [hep-ph]].


\bibitem{Karam:2015jta} 
  A.~Karam and K.~Tamvakis,
  Phys.\ Rev.\ D {\bf 92}, no. 7, 075010 (2015)
  doi:10.1103/PhysRevD.92.075010
  [arXiv:1508.03031 [hep-ph]].


\bibitem{DiazCruz:2010dc} 
  J.~L.~Diaz-Cruz and E.~Ma,
  Phys.\ Lett.\ B {\bf 695}, 264 (2011)
  doi:10.1016/j.physletb.2010.11.039
  [arXiv:1007.2631 [hep-ph]].


\bibitem{Barman:2017yzr} 
  B.~Barman, S.~Bhattacharya, S.~K.~Patra and J.~Chakrabortty,
  JCAP {\bf 1712}, 021 (2017)
  doi:10.1088/1475-7516/2017/12/021
  [arXiv:1704.04945 [hep-ph]].


\bibitem{Barman:2018esi} 
  B.~Barman, S.~Bhattacharya and M.~Zakeri,
  JCAP {\bf 1809}, 023 (2018)
  doi:10.1088/1475-7516/2018/09/023
  [arXiv:1806.01129 [hep-ph]].


\bibitem{Barman:2019lvm} 
  B.~Barman, S.~Bhattacharya and M.~Zakeri,
  JCAP {\bf 2002}, 029 (2020)
  doi:10.1088/1475-7516/2020/02/029
  [arXiv:1905.07236 [hep-ph]].


\bibitem{1712.08994} 
  E.~Ma,
  Phys.\ Lett.\ B {\bf 780}, 533 (2018)
  doi:10.1016/j.physletb.2018.03.053
  [arXiv:1712.08994 [hep-ph]].


\bibitem{Hill:2000mu} 
  C.~T.~Hill, S.~Pokorski and J.~Wang,
  Phys.\ Rev.\ D {\bf 64}, 105005 (2001)
  doi:10.1103/PhysRevD.64.105005
  [hep-th/0104035].


\bibitem{ArkaniHamed:2001ca} 
  N.~Arkani-Hamed, A.~G.~Cohen and H.~Georgi,
  Phys.\ Rev.\ Lett.\  {\bf 86}, 4757 (2001)
  doi:10.1103/PhysRevLett.86.4757
  [hep-th/0104005].


\bibitem{Georgi:1985hf} 
  H.~Georgi,
  Nucl.\ Phys.\ B {\bf 266}, 274 (1986).
  doi:10.1016/0550-3213(86)90092-1


\bibitem{Abe:2012hb} 
  T.~Abe, M.~Kakizaki, S.~Matsumoto and O.~Seto,
  Phys.\ Lett.\ B {\bf 713}, 211 (2012)
  doi:10.1016/j.physletb.2012.05.051
  [arXiv:1202.5902 [hep-ph]].


\bibitem{1202.5073} 
  K.~Hally, H.~E.~Logan and T.~Pilkington,
  Phys.\ Rev.\ D {\bf 85}, 095017 (2012)
  doi:10.1103/PhysRevD.85.095017
  [arXiv:1202.5073 [hep-ph]].


\bibitem{1906.05609} 
  G.~Aad {\it et al.} [ATLAS Collaboration],
  Phys.\ Rev.\ D {\bf 100}, no. 5, 052013 (2019)
  doi:10.1103/PhysRevD.100.052013
  [arXiv:1906.05609 [hep-ex]].


\bibitem{Sirunyan:2018mpc} 
  A.~M.~Sirunyan {\it et al.} [CMS Collaboration],
  JHEP {\bf 1806}, 128 (2018)
  doi:10.1007/JHEP06(2018)128
  [arXiv:1803.11133 [hep-ex]].


\bibitem{ATL-PHYS-PUB-2018-044} 
  The ATLAS collaboration [ATLAS Collaboration],
  ATL-PHYS-PUB-2018-044.


\bibitem{hep-ph/0405040} 
  R.~Barbieri, A.~Pomarol, R.~Rattazzi and A.~Strumia,
  Nucl.\ Phys.\ B {\bf 703}, 127 (2004)
  doi:10.1016/j.nuclphysb.2004.10.014
  [hep-ph/0405040].


\bibitem{1909.02845} 
  G.~Aad {\it et al.} [ATLAS Collaboration],
  Phys.\ Rev.\ D {\bf 101}, no. 1, 012002 (2020)
  doi:10.1103/PhysRevD.101.012002
  [arXiv:1909.02845 [hep-ex]].


\bibitem{hep-ph/9807565} 
  T.~Hahn and M.~Perez-Victoria,
  Comput.\ Phys.\ Commun.\  {\bf 118}, 153 (1999)
  doi:10.1016/S0010-4655(98)00173-8
  [hep-ph/9807565].


\bibitem{hep-ph/0512090} 
  M.~Cirelli, N.~Fornengo and A.~Strumia,
  Nucl.\ Phys.\ B {\bf 753}, 178 (2006)
  doi:10.1016/j.nuclphysb.2006.07.012
  [hep-ph/0512090].


\bibitem{1712.02118} 
  M.~Aaboud {\it et al.} [ATLAS Collaboration],
  JHEP {\bf 1806}, 022 (2018)
  doi:10.1007/JHEP06(2018)022
  [arXiv:1712.02118 [hep-ex]].


\bibitem{1801.03509} 
  G.~Bélanger, F.~Boudjema, A.~Goudelis, A.~Pukhov and B.~Zaldivar,
  Comput.\ Phys.\ Commun.\  {\bf 231}, 173 (2018)
  doi:10.1016/j.cpc.2018.04.027
  [arXiv:1801.03509 [hep-ph]].


\bibitem{Hisano:2004pv} 
  J.~Hisano, S.~Matsumoto, M.~M.~Nojiri and O.~Saito,
  Phys.\ Rev.\ D {\bf 71}, 015007 (2005)
  doi:10.1103/PhysRevD.71.015007
  [hep-ph/0407168].


\bibitem{Hisano:2010fy} 
  J.~Hisano, K.~Ishiwata and N.~Nagata,
  Phys.\ Lett.\ B {\bf 690}, 311 (2010)
  doi:10.1016/j.physletb.2010.05.047
  [arXiv:1004.4090 [hep-ph]].


\bibitem{Hisano:2015rsa} 
  J.~Hisano, K.~Ishiwata and N.~Nagata,
  JHEP {\bf 1506}, 097 (2015)
  doi:10.1007/JHEP06(2015)097
  [arXiv:1504.00915 [hep-ph]].


\bibitem{1310.1921} 
  A.~Alloul, N.~D.~Christensen, C.~Degrande, C.~Duhr and B.~Fuks,
  Comput.\ Phys.\ Commun.\  {\bf 185}, 2250 (2014)
  doi:10.1016/j.cpc.2014.04.012
  [arXiv:1310.1921 [hep-ph]].


\bibitem{Hisano:2006nn} 
  J.~Hisano, S.~Matsumoto, M.~Nagai, O.~Saito and M.~Senami,
  Phys.\ Lett.\ B {\bf 646}, 34 (2007)
  doi:10.1016/j.physletb.2007.01.012
  [hep-ph/0610249].


\bibitem{Cirelli:2005uq} 
  M.~Cirelli, N.~Fornengo and A.~Strumia,
  Nucl.\ Phys.\ B {\bf 753}, 178 (2006)
  doi:10.1016/j.nuclphysb.2006.07.012
  [hep-ph/0512090].


\bibitem{Aprile:2015uzo} 
  E.~Aprile {\it et al.} [XENON Collaboration],
  JCAP {\bf 1604}, 027 (2016)
  doi:10.1088/1475-7516/2016/04/027
  [arXiv:1512.07501 [physics.ins-det]].


\bibitem{hep-ph/0212022} 
  J.~Hisano, S.~Matsumoto and M.~M.~Nojiri,
  Phys.\ Rev.\ D {\bf 67}, 075014 (2003)
  doi:10.1103/PhysRevD.67.075014
  [hep-ph/0212022].


\bibitem{hep-ph/0307216} 
  J.~Hisano, S.~Matsumoto and M.~M.~Nojiri,
  Phys.\ Rev.\ Lett.\  {\bf 92}, 031303 (2004)
  doi:10.1103/PhysRevLett.92.031303
  [hep-ph/0307216].


\bibitem{hep-ph/0412403} 
  J.~Hisano, S.~Matsumoto, M.~M.~Nojiri and O.~Saito,
  Phys.\ Rev.\ D {\bf 71}, 063528 (2005)
  doi:10.1103/PhysRevD.71.063528
  [hep-ph/0412403].


\bibitem{0810.0713} 
  N.~Arkani-Hamed, D.~P.~Finkbeiner, T.~R.~Slatyer and N.~Weiner,
  Phys.\ Rev.\ D {\bf 79}, 015014 (2009)
  doi:10.1103/PhysRevD.79.015014
  [arXiv:0810.0713 [hep-ph]].


\bibitem{1603.01383} 
  K.~Blum, R.~Sato and T.~R.~Slatyer,
  JCAP {\bf 1606}, 021 (2016)
  doi:10.1088/1475-7516/2016/06/021
  [arXiv:1603.01383 [hep-ph]].
  
  
  

\end{thebibliography}
\end{document}